\documentclass[aps,prb,twocolumn,superscriptaddress]{revtex4-1}
\usepackage[version=3]{mhchem}  
\usepackage[warning,np]{numprint}
\usepackage{booktabs}
\npdecimalsign{\ensuremath{.}}
\npthousandsep{\,}
\npproductsign{\times}
\npfourdigitnosep

\newcommand{\cm}[1]{\np[cm^{-1}]{#1}}
\newcommand{\cmFWHM}[1]{\np[cm^{-1}\,FWHM]{#1}}

\usepackage{xcolor}   
\usepackage{booktabs} 
\usepackage{graphicx}

\begin{document}

\title{The spin-forbidden vacuum-ultraviolet absorption spectrum of \ce{{}^{14}N^{15}N}}
\author{A.~N. Heays}
\affiliation{Department of Physics and Astronomy, and LaserLaB, VU University, De Boelelaan 1081, 1081 HV Amsterdam, The Netherlands}
\affiliation{Leiden Observatory, Leiden University, P.O. Box 9513, 2300 RA Leiden, The Netherlands}
\affiliation{School of Earth \& Space Exploration, Arizona State University, Tempe, AZ USA}
\affiliation{NASA Astrobiology Institute, NASA Ames Research Center, Moffett Field, California, USA}
\affiliation{Corresponding author: aheays@asu.edu}
\author{B.~R. Lewis}
\affiliation{Research School of Physics and Engineering, The Australian National University, Canberra, Australian Capital Territory 0200, Australia}
\author{N.~de~Oliveira}
\affiliation{Synchrotron Soleil, Orme des Merisiers, St. Aubin, BP 48, 91192 Gif sur Yvette Cedex, France}
\author{W.~Ubachs}
\affiliation{Department of Physics and Astronomy, and LaserLaB, VU University, De Boelelaan 1081, 1081 HV Amsterdam, The Netherlands}

\begin{abstract}
  Photoabsorption spectra of \ce{{}^{14}N^{15}N} were recorded at high resolution with a vacuum-ultraviolet Fourier-Transform spectrometer fed by synchrotron radiation in the range 81 to 100 nm. The combination of high column density (\np[cm^{-2}]{3e17}) and low temperature (\np[K]{98}) allowed for the recording of weak spin-forbidden absorption bands exciting levels of triplet character. The triplet states borrow intensity from ${}^1\Pi_u$ states of Rydberg and valence character while causing their predissociation.
  New predissociation linewidths and molecular constants are obtained for the states $C\,{}^3\Pi_u(v=7,8,14,15,16,21)$, $G\,{}^3\Pi_u(v=0,1,4)$, and $F\,{}^3\Pi_u(v=0)$.
  The positions and widths of these levels are shown to be well-predicted by a coupled-Schr\"{o}dinger equation model with empirical parameters based on experimental data on \ce{{}^{14}N_2} and \ce{{}^{15}N_2} triplet levels.
\end{abstract}

\maketitle

\section{Introduction}
\label{sec:introduction}

The vacuum-ultraviolet (VUV) photoabsorption spectrum of molecular nitrogen is of importance to studies of terrestrial atmospheric physics,\cite{Bishop2007,Liu2009} planetary physics\cite{Lunine1999,Stevens2001,Lavvas2011} and astrophysics.\cite{Li2013,Li2014,Heays2014a}
It is also of fundamental interest in molecular physics in view of the benchmark perturbation phenomena occurring in this diatomic molecule.
Early investigations of the dipole-allowed absorption spectrum of N$_2$ revealed the perturbations between singlet states of valence and Rydberg character,\cite{Lefebvre1969,Dressler1969,Carroll1969} giving rise to intensity-interference effects in the vibronic spectrum, even extending into details of the rotational fine structure.\cite{Edwards1995,Vieitez2007} Interactions between the singlet manifolds and the triplet manifolds, induced by spin-orbit coupling, were found to cause strong predissociation effects in the excited states, however, in most cases, still allowing for resolved rotational structure.\cite{Ubachs1989}

Over the decades a large number of spectroscopic investigations of N$_2$ have been performed, focusing on the predissociation phenomenon, and using a variety of advanced techniques, ranging from tunable narrowband extreme ultraviolet (XUV) laser spectroscopy,\cite{Levelt1992,Ubachs2000,Sommavilla2006} picosecond pump-probe experiments,\cite{Ubachs2001,Sprengers2004a} electron-impact excitation,\cite{Ajello1989,Khakoo2008} synchrotron absorption studies,\cite{Huber1990,Stark1992,Stark2000,Stark2005} synchrotron fluorescence-excitation studies,\cite{Wu2012} detection of dissociation products in charge-exchange studies,\cite{Helm1993,Kamp1994a} as well as upon VUV-laser excitation,\cite{Song2016} and the spectroscopic study of infrared transitions within the triplet system.\cite{Kanamori1991,Hashimoto2006}

A framework of coupled-channel Schr\"{o}dinger equations (CSE) was developed\cite{Gibson1996,Haverd2005,Lewis2008b,Ndome2008,Heays2011_thesis} to provide a description of the absorption spectrum of molecular nitrogen, including Rydberg-valence electrostatic interactions and singlet-triplet spin-orbit interactions. Through this description, a comprehensive explanation is provided for the spectral perturbations as well as the predissociation phenomena.\cite{Lewis2005b,Lewis2008a,Lewis2008c}
Based on the CSE-framework, the full absorption spectrum and dissociation cross section of N$_2$ has been calculated,\cite{Li2013} as well as the electron-impact XUV emission spectrum.\cite{Heays2014b}
In principle, the CSE model is applicable to all isotopologues of nitrogen through mass scaling of the molecular parameters involved, and in some cases it has been applied to \ce{{}^{14}N^{15}N} and \ce{{}^{15}N_2},\cite{Sprengers2003,Sprengers2005b,Vieitez2008a} comparing with data from XUV laser excitation.
One application resulted in the calculation of the isotopically selective dissociation of nitrogen in the interstellar radiation field.\cite{Heays2014a}

The most complete investigation of the absorption spectrum of \ce{{}^{14}N^{15}N} was performed using VUV Fourier-Transform (FT) spectroscopy, analyzing in detail 25 singlet-singlet bands of this isotopologue\cite{Heays2011} and identifying perturbations in the level structure of singlet states caused by a parallel manifold of optically-forbidden states of triplet character.
The present study is an extension of this, whereby triplet states are observed directly by using an increased gas column density and by recording absorption at lower temperatures.
The motivation of the present work is to test the ability of a CSE model, developed by Lewis {\em et al.}\cite{Lewis2008b} with reference to the \ce{{}^{14}N_2} and \ce{{}^{15}N_2} isotopologues, to predict the triplet spectrum of \ce{{}^{14}N^{15}N}.
In the longer term, the new experimental data on the ${^3\,\Pi_u}$ states of \ce{{}^{14}N^{15}N} presented here,\cite{FN1} together with similar
planned measurements on \ce{{}^{15}N_2}, will enable the improvement and extension of the CSE model of N$_2$ predissociation, with 
particular application to studies of isotopic fractionation.

\section{Experimental details}
\label{sec:Experimental details}

High-resolution absorption spectra of \ce{{}^{14}N^{15}N} were recorded with the VUV-FT instrument installed at the DESIRS beam line of the SOLEIL synchrotron.\cite{Oliveira2011,Oliveira2016} This instrument is based upon a wave-front interferometer that allows the extension of the FT technique into the deep VUV. Spectral coverage is determined by the undulator synchrotron radiation source, delivering a bandwidth of about 5\,nm illuminating the FT instrument. The \ce{{}^{14}N^{15}N} gas flows continuously through a 100\,mm-long windowless absorption cell with capillaries at both ends, thus providing a condition of quasi-static gas pressure inside the cell.
The absolute frequency calibration is based on the intrinsically-linear frequency scale generated by the FT spectrometer, and from comparison of selected \ce{H2} and Ar\,I lines appearing in the spectrum with high-precision reference data.
The specific absorption lines used are Ar\,I at \cm{95399.833};\cite{Velchev1999} \ce{H2} at \np{99755.399}, \np{101668.268}, \np{101746.449}, and \cm{101825.295};\cite{Bailly2010} and \ce{H2} at \np{107114.785} and \cm{105549.599}.\cite{Philip2004b}
The estimated absolute calibration uncertainty is \cm{0.004}, but individual \ce{{}^{14}N^{15}N} line wavenumbers have greater uncertainties due to fitting errors of $\sim 0.02$\,cm$^{-1}$, or more for lines that are blended or saturated.

\begin{figure*}
  \centering
  \includegraphics{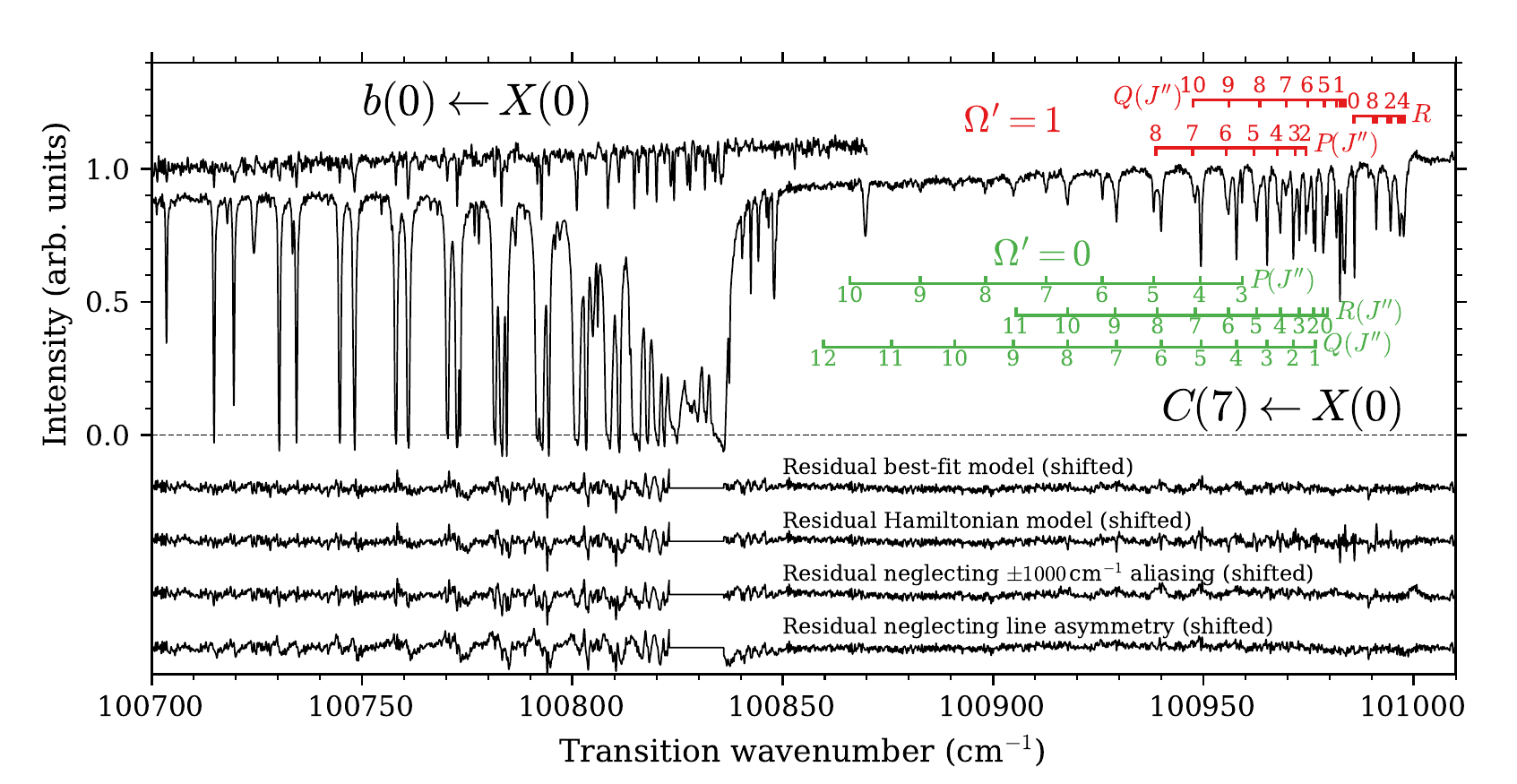}
  \caption{Absorption spectra showing $b(0)\leftarrow X(0)$ and $C(7)\leftarrow X(0)$. \emph{Upper spectrum:} Room-temperature low-column-density spectrum analyzed by Heays {\em et al.}\cite{Heays2011} \emph{Lower spectrum:} Liquid-\ce{N2}-cooled spectrum. Also indicated are the residual errors of models fitted to the lower spectrum under various assumptions discussed in the text.}
  \label{fig:C07_spectrum}
\end{figure*}

Our experimental and analysis procedures are similar to those previously adopted by Heays {\em et al.}\cite{Heays2011} except that measurements were performed with the cell externally cooled by a flowing liquid-nitrogen jacket, and employing $\sim 300\times$ greater column density.
The optically-allowed transitions to  ${}^1\Pi_u$ and ${}^1\Sigma^+_u$ states previously studied\cite{Heays2011} are then completely saturated in our spectra near their bandheads, but the targeted forbidden transitions exciting ${}^3\Pi_u$ levels are now measurable; e.g., in Fig.~\ref{fig:C07_spectrum} showing $b\,{}^1\Pi_u(v=0)\leftarrow X\,{}^1\Sigma^+_g(v=0)$ and $C\,{}^3\Pi_u(v=7)\leftarrow X\,{}^1\Sigma^+_g(v=0)$.
The main advantage of using cooled gas is the reduced spectral congestion from singlet-state absorption, due to fewer thermally-populated ground state rotational levels.
Additionally, the profiles of predissociation-broadened ${}{^3\Pi_u}\leftarrow X\,{^1\Sigma^+_g}$ bands are more revealing, with their oscillator strengths concentrated into $J''\lesssim 10$ transitions.

The population distribution of  \ce{{}^{14}N^{15}N} ground-state rotational levels was assumed to correspond to a mixture of room-temperature (295\,K) and cold gases, with the column density, $N_i$, of each, and the cold-gas rotational temperature determined by a fit to the unsaturated rotational transitions of the $b(0)\leftarrow X(0)$ (Fig.~\ref{fig:C07_spectrum}) and $b(1)\leftarrow X(0)$ bands.\cite{FN4}
The resulting best-fit values are $N_\text{room}=(2.2\pm 0.3)\times 10^{15}\,\text{cm}^{-2}$, $N_\text{cold}=(3.9\pm 0.6)\times 10^{17}\,\text{cm}^{-2}$, and $T_\text{cold}=98\pm 5$\,K.
The absolute $N_i$ calibration was made with reference to $b(0)\leftarrow X(0)$ and $b(1)\leftarrow X(0)$ $f$-values determined by Heays {\em et al.}\cite{Heays2011} and has an estimated 15\% uncertainty arising from a combination of the estimated uncertainty of Heays {\em et al.}\cite{Heays2011} and our own fitting errors.
The cooled-gas temperature uncertainty was estimated subjectively based on its range giving a good quality-of-fit to the $b(v)\leftarrow X(0)$ spectrum.
The assumed two-temperature model fits the spectrum quite well and we found no benefit from adopting more complicated temperature profiles.
Contributions to the experimental spectra from species other than the desired \ce{{}^{14}N^{15}N} were found, inevitably from
cold \ce{{}^{14}N2} and \ce{{}^{15}N2}, yielding, e.g., weak lines overlapping $b(0)\leftarrow X(0)$ in Fig.~\ref{fig:C07_spectrum}, and from room-temperature \ce{H2} contaminant, e.g., at \cm{100870} in Fig.~\ref{fig:C07_spectrum}.

The 98\,K Doppler width of \ce{{}^{14}N^{15}N} resonances amounts to \cmFWHM{0.14} (full-width half-maximum) at \cm{105000}. Spectra were recorded at a setting of the VUV-FT instrument giving rise to an unapodised sinc-shaped instrumental lineshape with width \cmFWHM{0.215}.
Several non-ideal instrumental effects were included in the modelling of the total instrumental broadening function.
Asymmetric lineshapes are evident in spectral regions near highly-saturated bandheads of ${}^1\Pi_u\leftarrow X(0)$ and ${}^1\Sigma^+_u\leftarrow X(0)$ transitions where the near-zero recorded intensity appears to locally interfere with the frequency-dependent FT phase correction.\cite{Learner1996,Brault1987}
This effect was modelled phenomenologically by convolution of the ideal instrumental function with the following:
\begin{equation}
  F_\text{asym}(\nu) =
  \begin{cases}
    1 & \text{if }\nu=0\\
    a/\nu & \text{otherwise} ,
  \end{cases}
\end{equation}
where $\nu$ is the wavenumber gap from line centre and the parameter $a$ was fitted to each spectral region studied.
An example of the degradation of the model fit to the measured spectra from neglecting this asymmetry is shown in Fig.~\ref{fig:C07_spectrum}, but this effect is significant only for those ${}^3\Pi_u\leftarrow X(0)$ bands overlapped by strong singlet-state excitation.

A second non-ideal instrumental effect appears as a weak aliasing of the experimental spectrum shifted by $\pm\cm{1000}$, where the aliasing is additive for one sign and subtractive for the other.
This is due to a periodic error in the stepping system of the FT translation\cite{Learner1996} and was handled by counteracting the aliasing through the addition and subtraction of $\pm$\cm{1000}-shifted copies of the experimental spectrum, with best-fit correction magnitude determined separately for each spectral-region studied.
Neglecting this aliasing introduces a significant model error near \cm{101000} in Fig.~\ref{fig:C07_spectrum}, e.g., which is due to negative aliasing of the strong $b(2)\leftarrow X(0)$ absorption at \cm{+1000}.
The magnitude of this correction is 1--3\% of the measured intensity and second-order corrections were deemed unnecessary.

Finally, an additional phenomenological Gaussian-shaped broadening was included in the instrumental broadening, with width  \cmFWHM{0.1}, as found necessary for this instrument by Heays {\em et al.}\cite{Heays2011}
Full information could not be retrieved from the experimental spectrum in heavily saturated regions, even with these corrections, as is evident in the residual error of the best-fit model shown in Fig.~\ref{fig:C07_spectrum} where the worst affected portions were neglected from model fitting (zero residual).

\section{Spectral analysis methods}
\label{sec:analysis}
\begin{figure}
  \centering
  \includegraphics{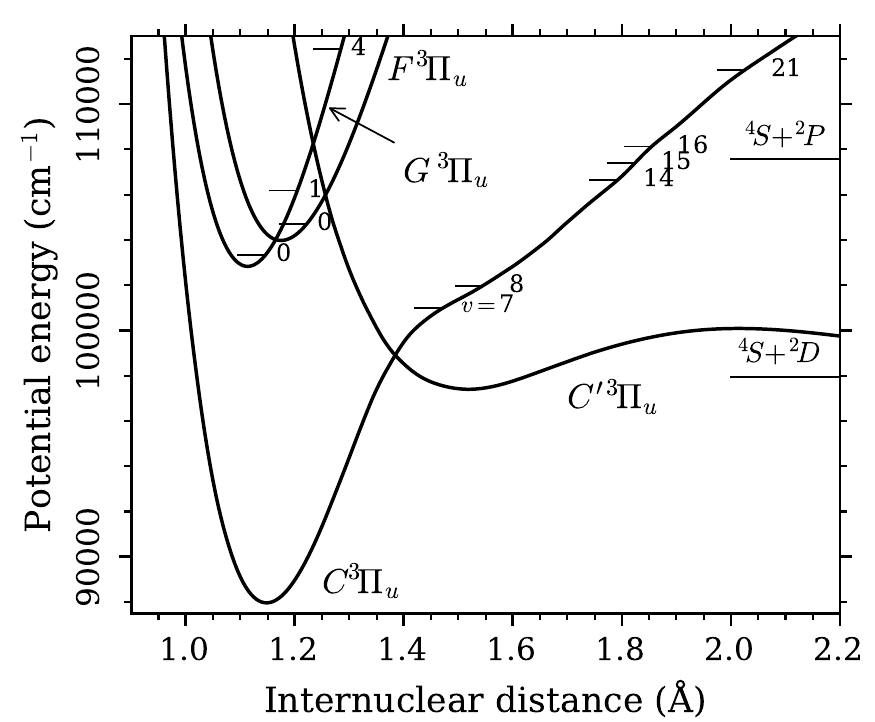}
  \caption{Potential-energy curves of ${}^3\Pi_u$ states (after Ref.~\onlinecite{Lewis2008b}), referenced to $X\,{}^1\Sigma^+_g(v=0,J=0)$ of \ce{{}^{14}N^{15}N} and with the newly-observed levels indicated.}
  \label{fig:potentials}
\end{figure}

Molecular parameters summarizing the observed $^3\Pi_u$ states were extracted by fitting model spectra to the experimental absorption measurements.
Each observed \ce{{}^{14}N^{15}N} line was assigned a transition wavenumber,
 $\nu_i$, line $f$-value according to the definition of Morton and Noreau, \cite{Morton1994} $f_i$, and predissociation linewidth, $\Gamma_i$.\cite{FN2}
Resulting temperature-dependent line cross sections were computed according to
\begin{equation}
  \label{eq:line cross section}
  \sigma_i(\nu;T) = \left(\frac{e^2\cdot 10^2}{4\epsilon_0m_ec^2}\right) f_i \alpha_{J_i''}(T) V\bigl(\nu-\nu_i,\Gamma_i,\Gamma_\text{D}(T)\bigr)
\end{equation}
where $\sigma_i$ is in cm$^2$, $T$ is the temperature in Kelvin, and $e$, $\epsilon_0$, $m_e$, and $c$ are the usual physical constants.
The parenthesised prefactor has the numerical value \np[cm]{8.853e-13}.
In Eq.~(\ref{eq:line cross section}), $V$ is a unit-area Voigt function centred on $\nu_i$ with Lorentzian width $\Gamma_i$ and Gaussian (Doppler) width $\Gamma_\text{D}(T)$; and $\alpha_{J_i''}(T)$ is the Boltzmann population of ground-state rotational level $J_i''$.

A total absorption opacity was constructed, allowing for absorption by \ce{{}^{14}N^{15}N} lines at 98\,K and 295\,K, together with contaminants:
\begin{multline}
 \label{eq:opacity}
  \tau(\nu) = \sum_i N_\text{98K}\sigma_{i}(\nu;\text{98K}) + N_\text{295K}\sigma_{i}(\nu;\text{295K}) \\
  + \sum_{j} \tau^\text{int}_{j}V(\nu-\nu_j,\Gamma_j,\Gamma_{\text{D}j}),
\end{multline}
using the \ce{{}^{14}N^{15}N} column densities discussed in Sec.~\ref{sec:Experimental details}.
The second summation in Eq.~(\ref{eq:opacity}) accounts for any \ce{{}^{14}N_2}, \ce{{}^{15}N_2} or \ce{H_2} contaminant lines using their fitted frequencies, integrated optical depths, $\tau^\text{int}_{j}$, and line shapes.

This opacity was converted into an absorption spectrum:\cite{FN3}
\begin{equation}
  \label{eq:absorption spectrum}
  I(\nu) = \bigl[ I_0(\nu)e^{-\tau(\nu)} \bigr] \ast F(\nu),
\end{equation}
for a direct pointwise comparison with the experimental spectrum.
Here, the incoming intensity of synchrotron radiation, $I_0(\nu)$, was fitted to a slowly-varying function of $\nu$; and the ideal absorption spectrum broadened by convolution with the instrumental function, $F(\nu)$, described in Sec.~\ref{sec:Experimental details}.

The $Q$-branch rotational transitions of strong and unblended ${}^3\Pi_u\leftarrow X\,{}^1\Sigma^+_g$ bands were modelled individually and their $P(J'+1)$ and $R(J'-1)$ transition in pairs, with a common upper-state term value and linewidth, different ground-state $X(v=0,J'')$ term values, and different oscillator strengths.
Ground-state energy levels were computed from the Dunham coefficients of Le~Roy {\em et al.}\cite{Leroy2006}

It was not possible to fit parameters to the individual rotational lines of the weak or broadened features observed experimentally.
Instead, smaller sets of parameters were optimized, controlling an effective Hamiltonian matrix representing all upper-state levels.
For this purpose, the  Hund's case-(a) Hamiltonian of Brown and Merer\cite{Brown1979} was adopted, fully specified in their Table~1 for the case of a ${}^3\Pi$ state, and a standard Hamiltonian was used to represent the observed ${}^1\Pi$ and ${}^1\Sigma$ levels:
\begin{align}
  T_{{}^1\Sigma} &= T + BJ(J+1) - D\left[J(J+1)\right]^2, \\
\text{and }  T_{{}^1\Pi} &= T + BJ(J+1) - D\left[J(J+1)\right]^2 + qJ(J+1),
\end{align}
where the parameter $q$ is only included for $f$-parity levels.
The spin-orbit mixing of case-(a) ${}^3\Pi_u$ and ${}^1\Pi_u$ states is restricted to their $\Omega=1$ levels of common $e/f$ parity and is modelled with an interaction matrix element $\xi$, treated as a free parameter for each pair of interacting electronic-vibrational levels.
The mixed-level energies and mixing-coefficients are then computed by diagonalizing the full Hamiltonian matrix of ${}^3\Pi_{u\Omega}$ and ${}^1\Pi_{u1}$ levels.

The absorption strength of ${}^3\Pi_u\leftarrow X\,{}^1\Sigma^+_g(0)$ transitions is borrowed from nearby ${}^1\Pi_u\leftarrow X\,{}^1\Sigma^+_g(0)$ transitions through their mutual spin-orbit coupling.\cite{Lewis2005a}
We model the rotational line strengths of the ${}^1\Pi_u\leftarrow X\,{}^1\Sigma^+_g(0)$ transitions using an empirical electronic-vibrational transition moment and H\"onl-London rotational-linestrength factors.\cite{Lefebvre}
Mixing coefficients from the Hamiltonian-matrix diagonalization are then used to generate the full rotational structure of the observed mixed ${}^3\Pi_u \sim {}^1\Pi_u\leftarrow X\,{}^1\Sigma^+_g$ spectrum.

$\Omega$ labels are assigned to the observed ${}^3\Pi_u$ levels forming continuous $e/f$-parity rotational progressions, based on their dominant character at low $J$.
The relative contribution of nominal $\Omega$-levels to the observed absorption depends on their admixture of pure $\Omega=1$ character and is controlled by $J$ and the diagonal spin-orbit constant $A$.
Sufficiently large-$A$ and low-$J$ states are of nearly unmixed $\Omega$ character and transitions to the central $\Omega=1$ levels appear strongly.
For increasing $J$, these transitions weaken and the higher- and lower-energy triplet levels are more readily excited.
An example of this phenomenon is given in a detailed study on the $E{}^1\Pi \sim k{}^3\Pi$ interaction in isotopologues of CO.\cite{Cacciani1995}
Other level and intensity patterns may occur where the magnitudes of $A$ and the rotational constant $B$ are comparable,\cite{Brown1979} e.g., as found for the $C\,{}^3\Pi_u(v=7)$ level.

  The spin-orbit borrowing of absorption intensity is reciprocated by a sharing of dissociation proclivity. However, while the source of the intensity is always the
dipole-allowed ${}^1\Pi_u \leftarrow X(0)$ transition, both the ${}^3\Pi_u$ and ${}^1\Pi_u$ levels have nonzero inherent predissociation widths, prior to mixing,
which are derived from different multichannel dissociation pathways, although ultimately involving the $C^{\prime}\,{^3\Pi_u}$ valence state. This can lead to interference effects and $J$-dependent linewidths.
Here, we adopt a hybrid procedure to fit the experimental ${}^3\Pi_u \leftarrow X(0)$ band cross sections, 
empirically modelling the dissociation linewidth of ${}^3\Pi_u$ effective-Hamiltonian levels by means of an imaginary energy, $i\Gamma$,\cite{Cacciani1995}
mixing the ${}^3\Pi_u$ and ${}^1\Pi_u$ levels with the $\xi$ parameter, and using the results of previous work to treat the ${}^1\Pi_u \leftarrow X(0)$ transition.\cite{Heays2011} 
This approach is necessary since, as implied above, it is not usual to find simultaneous consistency between the two-state modelled linewidths and the corresponding intensity sharing, due to the inherent multilevel nature of the interaction. 
It should be noted,
however, that the independently-fitted ${}^3\Pi_u$ experimental linewidths determined here are meaningful and very suitable for comparison with the CSE computations.

\section{CSE modelling}

\begin{table*}
  \begin{minipage}{\textwidth}
  \caption{Experimental molecular constants of \ce{{}^{14}N^{15}N} ${}^3\Pi_u$ levels and those computed from the CSE model.\protect\footnote{All in \cm{} apart from $\Gamma$ (cm$^{-1}$\,FWHM) and $f_v$ (dimensionless). Estimated $1\sigma$ uncertainties are given in parentheses in units of the least-significant digit.}}
  \label{tab:fitted_constants}
  \begin{tabular}{lllllllll}
    \hline
    Level&\multicolumn{1}{c}{$T_{00}$\protect\footnote{$\Omega=1$, $J=0$ term origin, from adding $(2B-4\lambda/3-2\gamma)$ to the Ref.~\protect\onlinecite{Brown1979} Hamiltonian term, referenced to an $X(v=0,J=0)$ energy zero.}}&\multicolumn{1}{c}{$T_{00}^\text{CSE}$}&\multicolumn{1}{c}{$B$\protect\footnote{Entries without uncertainties are assumed values.}}&\multicolumn{1}{c}{$B^\text{CSE}$}&\multicolumn{1}{c}{$A$\footnotemark[3]}&\multicolumn{1}{c}{$\Gamma$}&\multicolumn{1}{c}{$\Gamma^\text{CSE}$}&\multicolumn{1}{c}{$f_v\times 10^4$\footnote{A 15\% uncertainty is assigned to $f$-values of well-resolved bands, corresponding to the overall column-density calibration uncertainty. Larger uncertainties are estimated for poorly resolved or fitted bands. }}\\
    \hline

$C(6)$
 & $$               & $\phantom{1}99\,733.2$ & $$           & $1.110$    & $$           & $$           & $0.33$     & $$           \\
$C(7)$\footnote{Results of a two-state deperturbation with $b(0)$ with additional parameters: $\lambda_{C(7)}=1.37(3)$, $\gamma_{C(7)}=-0.007(1)$, and $o_{C(7)}=0.695(8)$\,cm$^{-1}$. $T_{b(0)}=100832.33(4)$, $B_{b(0)}=1.40074(6)$, $D_{b(0)}=2.62(2)\times 10^{-5}$ and $\xi_{C(7)\sim b(0)}=16.5(2)$\,cm$^{-1}$.}
 & $100\,979.9(1)$  & $100\,981.8$  & $1.3613(3)$  & $1.364$    & $3.69(2)$    & $0.11(2)$\footnote{Width at $J=1$.} & $0.09$     & $0.24(4)$       \\
$C(8)$\footnote{Results of a two-state deperturbation with $b(2)$ with additional parameters: $T_{b(2)}=102138.71(1)$, $B_{b(2)}=1.34226(6)$, $D_{b(2)}=1.64(2)\times 10^{-5}$ and $\xi_{C(8)\sim b(2)}=17(1)$\,cm$^{-1}$.}
 & $101\,980(5)$    & $101\,970.7$  & $1.3$        & $1.279$    & $2$        & $60(15)$     & $55$       & $1.9(4)$        \\
$C(9)$
 & $$               & $102\,781$  & $$           & $1.31$    & $$           & $$           & $142$      & $$           \\
$C(10)$
 & $$               & $103\,543$  & $$           & $1.32$    & $$           & $$           & $167$      & $$           \\
$C(11)$
 & $$               & $104\,307$  & $$           & $1.24$    & $$           & $$           & $111$      & $$           \\
$C(12)$
 & $$               & $105\,090.3$  & $$           & $1.194$    & $$           & $$           & $56$       & $$           \\
$C(13)$
 & $$               & $105\,870.3$  & $$           & $1.150$    & $$           & $$           & $21$       & $$           \\
$C(14)$
 & $106\,639.4(3)$  & $106\,639.6$  & $1.18(2)$    & $1.151$    & $-14$        & $0.61(3)$    & $0.62$     & $0.09(3)$      \\
$C(15)$\footnote{Results of a two-state deperturbation with $b(9)$ with additional parameters: $\lambda_{C(15)}=0.474(7)$, $o_{C(15)}=0.490(8)$, $T_{b(9)}=107644.87(1)$, $B_{b(9)}=1.2000(1)$, $D_{b(9)}=2.01(6)\times 10^{-5}$, $H_{b(9)}=3.35(8)\times 10^{-8}$, $q_{b(9)}=2.31(2)\times 10^{-5}$ and $\xi_{C(15)\sim b(9)}=6.5(1)$\,cm$^{-1}$.}
 & $107\,384.86(1)$ & $107\,380.5$  & $1.1052(2)$  & $1.101$    & $-13.03(1)$  & $0.02(1)$\footnotemark[6] & $0.00$     & $0.067(19)$       \\
$C(16)$\footnote{Results of a two-state deperturbation with $b(10)$ with additional parameters: $T_{b(10)}=108262.4(1)$, $B_{b(10)}=1.1686(1)$, $D_{b(10)}=1.54(3)\times 10^{-5}$ and $\xi_{C(16)\sim b(10)}=8(1)$\,cm$^{-1}$.}
 & $108\,130(4)$    & $108\,130.0$  & $1.27(5)$    & $1.143$    & $-14$        & $22(2)$      & $16$       & $0.41(6)$       \\
$C(17)$
 & $$               & $108\,852.6$  & $$           & $1.085$    & $$           & $$           & $59$       & $$           \\
$C(18)$
 & $$               & $109\,543.6$  & $$           & $1.050$    & $$           & $$           & $30$       & $$           \\
$C(19)$
 & $$               & $110\,211.7$  & $$           & $1.049$    & $$           & $$           & $69$       & $$           \\
$C(20)$
 & $$               & $110\,820$\footnote{Estimated value from closely coupled $C(20)$ and $G(3)$ resonances.}  & $$           & $1.07$\footnotemark[10]    & $$           & $$           & $45$\footnotemark[10]       & $$           \\
$C(21)$
& $111\,503(20)$   & $111\,477.6$  & $1.0$        & $1.009$   & $-18$        & $64(15)$     & $78$       & $4.2(11)$        \\
$C(22)$
 & $$               & $112\,081.1$  & $$           & $0.953$   & $$           & $$           & $72$       & $$           \\
$G(0)$\footnote{Results of a two-state deperturbation with $b(4)$ with additional parameters: $q_{G(0)}=-0.037(2)$, $T_{b(4)}=103517.2(3)$, $B_{b(4)}=1.374(4)$, $D_{b(4)}=5.1(15)\times10^{-5}$, $H_{b(4)}=1.1(15)\times10^{-8}$, $q_{b(4)}=1.4(5)\times 10^{-4}$, and $\xi_{G(0)\sim b(4)} = 1.8(7)$\,cm$^{-1}$.}
 & $103\,320.4(1)$  & $103\,320.8$  & $1.804(3)$   & $1.822$    & $-8.0(2)$    & $0.11(1)$    & $0.09$    & $0.18(3)$       \\
$G(1)$
 & $106\,184(20)$   & $106\,139.6$  & $1.7(4)$     & $1.780$    & $-8$       & $19(8)$      & $9.8$      & $0.12(3)$       \\
$G(2)$
 & $$               & $107\,961.3$  & $$           & $1.713$    & $$           & $$           & $68$       & $$           \\
$G(3)$
 & $$               & $110\,740$\footnotemark[10]  & $$           & $1.77\footnotemark[10]$    & $$           & $$           & $250$\footnotemark[10]      & $$           \\
$G(4)$
 & $112\,440(5)$    & $112\,519.9$  & $1.8(2)$     & $1.699$    & $-8$       & $32(5)$      & $24$      & $0.57(9)$       \\
$F(0)$\footnote{Additional parameter: $\lambda = 0.28(8)$\,cm$^{-1}$.}
 & $104\,718.42(7)$ & $104\,717.5$  & $1.75(2)$    & $1.747$    & $-21.64(7)$  & $0.4(1)$     & $0.44$    & $0.074(11)$      \\
$F(1)$
 & $$               & $106\,550.4$  & $$           & $1.654$    & $$           & $$           & $16$      & $$           \\
$F(2)$
 & $$               & $108\,608$  & $$           & $1.69$    & $$           & $$           & $140$      & $$           \\
$F(3)$
 & $$               & $109\,870$  & $$           & $1.74$    & $$           & $$           & $167$      & $$           \\
$F(4)$
 & $$               & $111\,727.2$  & $$           & $1.682$    & $$           & $$           & $28$       & $$           \\

    \hline
  \end{tabular}
  \end{minipage}
\end{table*}

A semi-empirical CSE model of the interacting $C$, $C'$, $F$, and $G\,{}^3\Pi_u$ states of N$_2$ was developed by Lewis {\em et al.}\cite{Lewis2008b} in order to reproduce the energy levels, rotational constants, and predissociation broadening of all experimentally observed levels.
This data set consisted of \ce{{}^{14}N2} levels between \np{88900} and \cm{111000}; i.e., as high as $v=3$ for the $F\,{}^3\Pi_u$ and $G\,{}^3\Pi_u$ states, and $v=18$ for $C\,{}^3\Pi_u$; together with a few additional levels of \ce{{}^{15}N2}.
The parameters of their model are the diabatic potential-energy curves, $V_i(R)$, of the ${}^3\Pi_{u1}$ states, reproduced in Fig.~\ref{fig:potentials}, and the electrostatic interaction energies mixing them.
Together, the parameters form an $R$-dependent potential matrix:
\begin{equation}
  \label{eq:potential matrix}
  {\bf V}(R) = \begin{pmatrix}
   V_C(R)& 798    & 0     & 0   \\
   798   & V_{C'}(R)& 315   & 1175\\
   0     & 315    & V_F(R)& 1115\\
   0     & 1175   & 1115  &V_G(R)
  \end{pmatrix}
,
\end{equation}
where $R$ is the internuclear distance, and the off-diagonal interaction energies are in \cm{}.
Coupled radial vector wavefunctions, ${\boldsymbol\chi}(R)$, are solutions to a coupled-channel Schr\"odinger equation:
\begin{equation}
\frac{d^2}{dR^2}{\boldsymbol\chi}(R) = \frac{-2\mu}{\hbar^2}{\boldsymbol\chi}(R)\left[E-{\bf V}(R)\right],
\label{eq:coupled radial equation}
\end{equation}
where $\mu$ is the molecular reduced mass.
These wavefunctions exist for all energies, $E$, lying above the dissociation energy of $C'\,{}^3\Pi_u$,
which provides the predissociation mechanism.
Detailed descriptions of the CSE method applied to the predissociative coupling of bound and unbound states, and a method for solving Eq.~(\ref{eq:coupled radial equation}), have been given previously.\cite{Mies1980a,Torop1987,Johnson1978}

The model of Lewis {\em et al.}\cite{Lewis2008b} is restricted to ${}^3\Pi_{u\Omega=1}$ states and considers neither the fine-structure splittings leading to separate $\Omega=0$ and 2 states, nor spin-orbit interactions involving ${}^1\Pi_u$ levels. Some of these details have been considered elsewhere.\cite{Lewis2005a,Heays2011_thesis,Heays2011}

The potential-energy curves and interaction energies in Eq.~(\ref{eq:potential matrix}) were optimised by Lewis {\em et al.}\cite{Lewis2008b} to best match the available \ce{{}^{14}N2} and \ce{{}^{15}N2} level data, but are independent of nuclear mass and also applicable to \ce{{}^{14}N{}^{15}N}.
We utilise and test this mass-independence by adopting the CSE model of Lewis {\em et al.}\cite{Lewis2008b} unchanged and computing its predictions for \ce{{}^{14}N{}^{15}N} by substitution of the correct reduced mass in Eq.~(\ref{eq:coupled radial equation}).

Elements of the vector wavefunction in Eq.~(\ref{eq:coupled radial equation}),
\begin{equation}
  {\boldsymbol\chi}(R;E) =
  \begin{pmatrix}
    \chi_C(R;E) \\ \chi_{C'}(R;E) \\ \chi_F(R;E) \\ \chi_G(R;E)
  \end{pmatrix}
,
\end{equation}
constitute $R$-dependent mixing coefficients of the four coupled electronic states.
Radially integrating the bound-state coefficients gives $E$-dependent overall mixing coefficients,
\begin{equation}
  \label{eq:CSE resonances}
  c_i(E) = \int_0^{R_\text{max}} \left|\chi_i(R;E)\right|^2\,dR,
\end{equation}
with $i=C$, $G$, or $F$, that show resonant structure near the vibrational eigenenergies that would occur if these states were not coupled to the $C'$ continuum and truly bound.
The central energies and widths of these coupled-state resonances are then directly comparable with the experimentally-determined level energies and linewidths.
We fit Fano lineshapes to the $c_i(E)$ spectrum in order to make this comparison.
The choice of the outer integration limit, $R_\text{max}$, in Eq.~(\ref{eq:CSE resonances}) is not significant as long as it extends beyond the classically-allowed region of all bound states.

Computed level energies, predissociation linewidths, and rotational constants are given in Table~\ref{tab:fitted_constants}.
The potential-energy curves in Eq.~(\ref{eq:potential matrix}) correspond to virtual $J=0$ levels, and the rotational constant is computed from the difference between the $J=5$ and $J=0$ CSE level energies, i.e., $B=(T_5-T_0)/30$,
where the rotationally-excited level is computed using a potential matrix with diagonal elements modified according to:
\begin{equation}
  V_{ii}(R;J) = V_{ii}(R;J=0) + \frac{\hbar^2 J(J+1)}{2\mu R^2},
\end{equation}
with $J=5$.

\section{Results and discussion}

In the present study, \ce{{}^{14}N^{15}N} states with excitation energies between $100\,900$ and $112\,600$\,cm$^{-1}$ are investigated under high column-density conditions to reveal the forbidden $^3\Pi_u\leftarrow X\,{}^1\Sigma^+_g$ spectrum.
In the following subsections, we discuss observed absorption bands terminating on $C\,{}^3\Pi_u(v=7,8,14-16,21)$, $G\,{}^3\Pi_u(v=0-1,4)$, and $F\,{}^3\Pi_u(v=0)$.

Molecular constants determined for the observed $^3\Pi_u$ levels are listed in Table~\ref{tab:fitted_constants} and line lists for all observed transitions are provided in the Supplementary Material.
The band oscillator strengths, $f_v$, in Table~\ref{tab:fitted_constants}, are computed by summing the $f$-values of all rotational transitions weighted by their lower-level thermal populations, and are then only strictly applicable at 98\,K.
The listed term origins, $T_{00}$, refer to virtual $\Omega=1$, $J=0$ triplet-state levels and differ from the Hamiltonian terms of Ref.~\protect\onlinecite{Brown1979} as described in the table footnotes.

\subsection{$C\,{}^3\Pi_u(v=7)$}
\label{sec:C07}

We assign a well-resolved band appearing in our spectrum near \cm{101000} to $C(7)\leftarrow X(0)$, following its observed location in \ce{{}^{14}N_2} and \ce{{}^{15}N_2}.\citep{Lewis2008a}
The spectrum is plotted in Fig.~\ref{fig:C07_spectrum} with lines assigned to six rotational branches accessing $e$- and $f$-parity levels of the $\Omega=0$ and 1 substates.
The band occurs \cm{150} higher in energy than the bandhead of $b(0)\leftarrow X(0)$, no doubt the principal perturbation partner from which $C(7)-X(0)$ borrows its absorption intensity.
The observed $C(7)\leftarrow X(0)$ lines are less predissociation broadened than found in other isotopologues where they are entirely rotationally blended.\citep{Lewis2008a}

\begin{figure}
  \centering
  \includegraphics{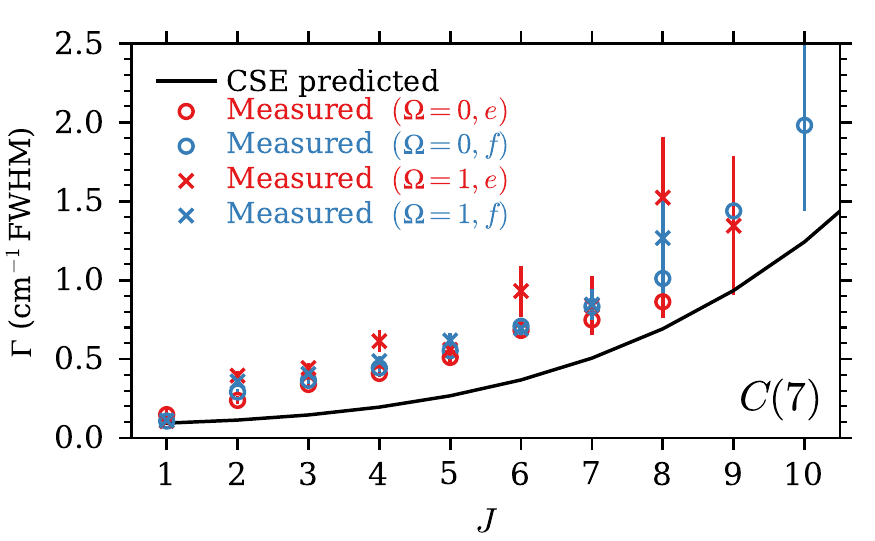}
  \caption{Rotational variation of predissociation linewidth for $C(7)$. Symbols: Experimental, this work. Curve: CSE computations using the
model of Ref.~\onlinecite{Lewis2008a}.} 
\label{fig:C07 widths}
\end{figure}
The observed spectrum was fitted to an effective Hamiltonian model for $C(7)$ and $b(0)$, coupled by a spin-orbit interaction.
The fitted parameters for $C(7)$ are listed in Table~\ref{tab:fitted_constants} and include a nonzero $\Lambda$-doubling parameter, $o$,\cite{Brown1979} to account for $e$- and $f$-parity splitting up to \cm{0.5}.
The interaction energy, chosen to reproduce the observed $C(7) \leftarrow X(0)$ intensity borrowing, does not provide a satisfactory value for the $b(0)$ predissociation broadening, underestimating this by a factor of 10 for low-$J$ levels when compared with the measured and calculated widths of Heays {\em et al.}\cite{Heays2011}
An initial model which included an empirical $J$-dependent linewidth for $b(0)$, following Heays {\em et al.},\cite{Heays2011} was used for the Hamiltonian model fit shown in Fig.~\ref{fig:C07_spectrum}.
However, this model does not fully explain the observed lineshapes of the $C(7)\leftarrow X(0)$ transitions.
In the finally adopted model, the term energy and linewidth for each $C(7)$ rotational level were allowed to vary individually, while retaining the Hamiltonian-derived line strengths.
The corresponding best-fit shown in Fig.~\ref{fig:C07_spectrum} is then quite satisfactory and the resulting $J$-dependent $C(7)$ predissociation widths are plotted in Fig.~\ref{fig:C07 widths}.

The deduced band origins and rotational constants, $T$ and $B$, respectively, are appropriately intermediate to those determined by Lewis {\em et al.}\cite{Lewis2008a} in observations of $C(7)$ in  \ce{{}^{14}N_2} and \ce{{}^{15}N_2}.
However, the \ce{{}^{14}N^{15}N} diagonal spin-orbit constant, $A=3.69\pm 0.01$ cm$^{-1}$, is found to be significantly smaller than the rough \cm{11} value found previously for both homonuclear species, although there is significant uncertainty in these previous estimates which were derived from rotationally-unresolved spectra.
A small spin-orbit constant is consistent with the known strong internuclear-distance-dependence in the principal configuration of $C\,{}^3\Pi_u$,\citep{Ndome2008} leading to a zero-crossing of $A$ as a function of vibrational level near $v=7$.
The energy ordering of $C(7)$'s pure Hund's case (a) $\Omega$-substates is affected by the smallness of $A$ and has the unusual sequence $\Omega=0$, 2, and 1.
Spin-rotation coupling thoroughly mixes the pure case (a) levels, even by $J=4$, but with near-zero admixture of pure $\Omega=1$ into the series of intermediate levels nominally-labelled $\Omega=2$.
No transitions to this $\Omega=2$ series are then observed in our spectrum because all intensity of forbidden absorption into $C\,{}^3\Pi_u$ arises from spin-orbit mixing of its pure $\Omega=1$ levels with $b\,{}^1\Pi_u$.


\subsection{$C\,{}^3\Pi_u(v=8)$}
\label{sec:C08}

\begin{figure}
  \centering
  \includegraphics{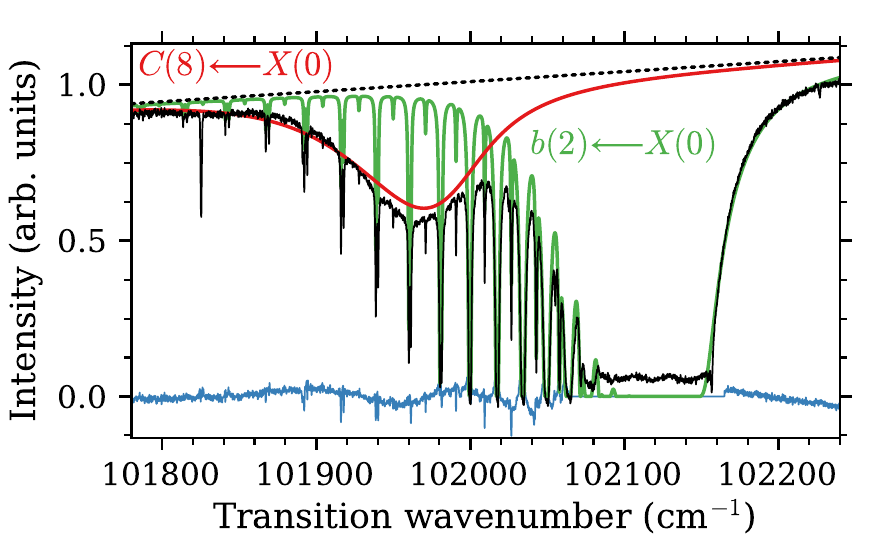}
  \caption{\emph{Black curve:} Absorption spectrum showing $C(8)\leftarrow X(0)$ and $b(2)\leftarrow X(0)$. \emph{Red curve:} $C(8)\leftarrow X(0)$ absorption computed from two-level deperturbation model.  \emph{Green curve:} $b(2)\leftarrow X(0)$ model absorption. \emph{Blue trace:} Residual error of the deperturbation model. \emph{Dotted line:} The frequency-dependent synchrotron radiation intensity.}
  \label{fig:C08_spectrum}
\end{figure}

Absorption due to $C(8)\leftarrow X(0)$ was observed in \ce{{}^{14}N_2} by Lewis {\em et al.}\cite{Lewis2008a} and we find a similar feature in our \ce{{}^{14}N ^{15}N} spectrum, centred at \cm{101950} and overlapped by $b(2)\leftarrow X(0)$ absorption lines, as shown in Fig.~\ref{fig:C08_spectrum}.
The rotational structure of $C(8)\leftarrow X(0)$ is completely blended and the red-degraded bandhead of $b(2)\leftarrow X(0)$ is totally saturated.
Apparent nonzero intensity in the bandhead region is due to a locally-unreliable phase correction of the measured interferogram where the true intensity is zero.
Furthermore, residual room-temperature \ce{{}^{14}N^{15}N} gas outside the cooled sample chamber contributes to the high-$J''$ lines at energies below \cm{101920}, and contaminating \ce{H2} lines appear in a few locations in the spectrum, e.g., at  \cm{101825}.

We constructed a local-perturbation model consisting of $C(8)$, $b(2)$, and their spin-orbit interaction, with fitted parameters given in Table~\ref{tab:fitted_constants}.
The resulting modelled and experimental absorption spectra are compared in Fig.~\ref{fig:C08_spectrum}.
This model was optimised with concurrent comparison to the unsaturated spectrum recorded by Heays {\em et al.},\cite{Heays2011} in order to better fit the $b(2)\leftarrow X(0)$ bandhead.

Only a few Hamiltonian parameters could be independently fitted to the unresolved $C(8)\leftarrow X(0)$ absorption and we therefore fixed the $C(8)$ diagonal spin-orbit constant to $A=$ \cm{2}, consistent with the values calculated for \ce{{}^{14}N_2} and \ce{{}^{15}N_2} by Ndome {\em et al.},\cite{Ndome2008} although the model fit is insensitive to any reasonably-small choice of $A$.
We also assume no $C(8)$ centrifugal distortion and $J$- and $\Omega$-independent predissociation broadening.
A freely-varied rotational constant $B=\cm{1.44}$ was found, but this parameter is only weakly constrained by the experimental spectrum and ultimately a fixed value $B=\cm{1.3}$ was assumed, given the values computed for \ce{{}^{14}N_2} and \ce{{}^{15}N_2} by Lewis {\em et al.}\cite{Lewis2008b}
The fitted predissociation linewidth of $C(8)$ is $\Gamma=60\pm15$\,cm$^{-1}$\,FWHM, with the uncertainty estimated by considering a range of values that plausibly fit the experimental spectrum.
Similarly subjective uncertainties were also evaluated for the term origin of $C(8)$ and the $C(8) \sim b(2)$ spin-orbit interaction energy.

The interaction between $b(2)$ and $C(8)$, optimised to best match the intensity borrowed by $C(8)\leftarrow X(0)$, also results in the correct \cmFWHM{0.6} line broadening for transitions to low-$J$ $b(2)$ levels.
Higher-$J$ $b(2)$ levels are known to have decreasing predissociation widths,\cite{Heays2011} but this effect is not reproduced by our two-level model.
The imperfect quality-of-fit of the modelled $C(8)\leftarrow X(0)$ band profile evident in Fig.~\ref{fig:C08_spectrum} is another indication that more than two interacting states are contributing to the observed $C(8)$ level.

\subsection{$C\,{}^3\Pi_u(v=14)$}

\begin{figure}
  \centering
  \includegraphics{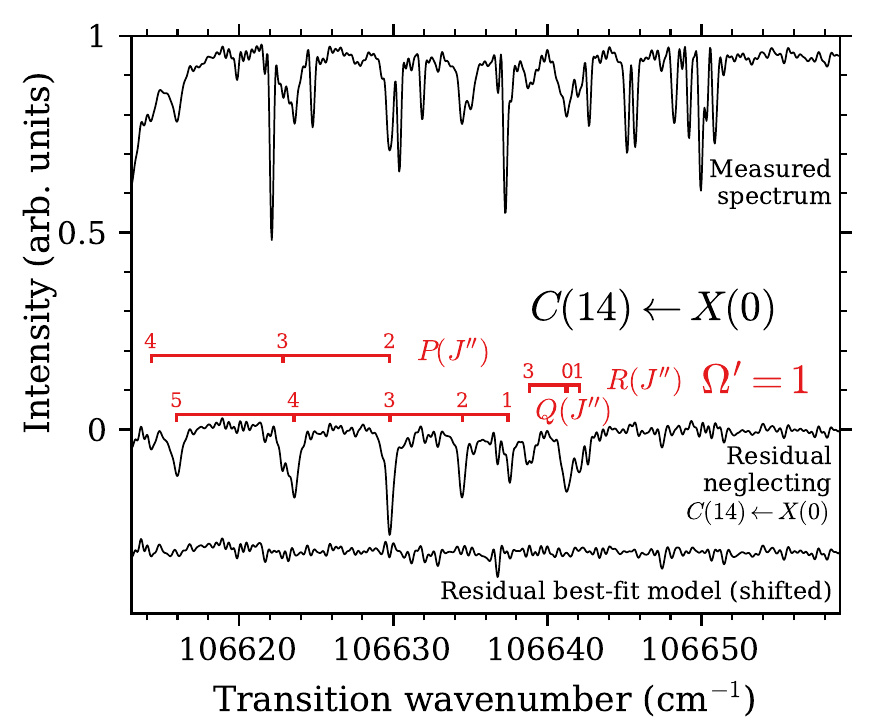}
  \caption{Spectrum showing $C(14)\leftarrow X(0)$ absorption and overlapping \ce{{}^{14}N{}^{15}N} and \ce{{}^{14}N_2} contamination. The residual error after modelling this spectrum is shown, and for a model neglecting $C(14)\leftarrow X(0)$, to highlight its rotational structure.}
  \label{fig:C14_spectrum}
\end{figure}
\begin{figure}
  \centering
  \includegraphics{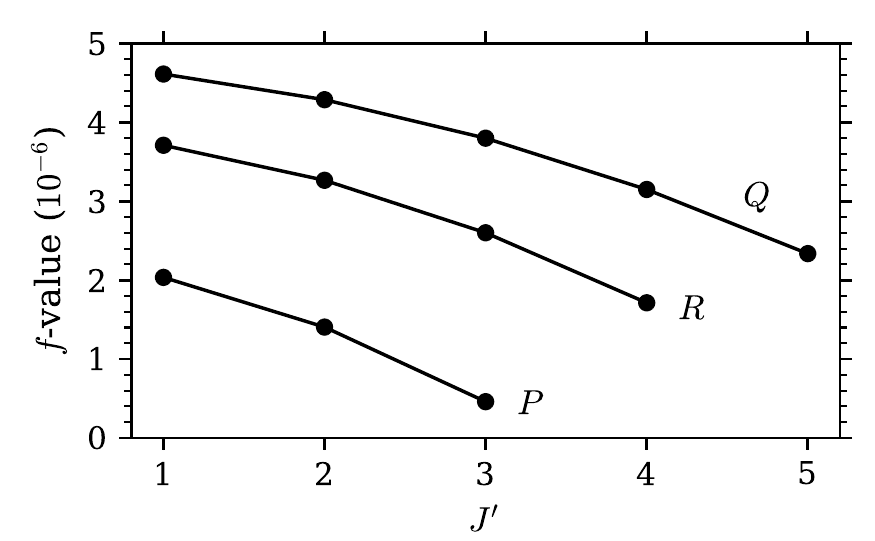}
  \caption{Experimental effective band $f$-values from observed branches of $C(v=14,\Omega=1)\leftarrow X(0)$, constrained to formulae: $f_v(J') = f_v(0)+{\rm const.}\times J'(J'+1)$.}
  \label{fig:C14_fvalues}
\end{figure}

Figure \ref{fig:C14_spectrum} shows a spectrum of $C(14) \leftarrow X(0)$ and overlying lines of \ce{{}^{14}N^{15}N} $b(8)\leftarrow X(0)$ and \ce{^{14}N_2} $b'(4)\leftarrow X(0)$, with steeply-rising absorption at lower energy due to the \ce{{}^{14}N^{15}N} bandhead of $b'(4)\leftarrow X(0)$.
After accounting for this contamination, $P$- and $R$-branch lines from $C(14) \leftarrow X(0)$, having $J'<4$ and the correct energy combination differences to originate from common $e$-parity upper levels, are evident in the spectrum.
Additional lines are observed from the $Q$-branch connecting with the corresponding $f$-levels, with $\Lambda$-doubling up to \cm{0.1}.
A common linewidth was fitted to all of these transitions, and $f$-values with assumed linear-in-$J'(J'+1)$ dependences fitted separately to each branch are plotted in Fig.~\ref{fig:C14_fvalues}.

We assigned the observed absorption to $C(14)\leftarrow X(0)$ due to its appearance midway between the corresponding band origins in \ce{^{14}N_2} and \ce{^{15}N_2}.\cite{Lewis2008b}
Absorption is only observed to one $\Omega$-component, with $\Omega=2$ ruled out by the observance of a $J=1$ level.
The strongest low-rotation transitions will occur for $\Omega'=1$ because of a direct admixture of this state with the ${}^1\Pi_u$ manifold, but with spin-rotation mixing sharing this intensity with nominal $\Omega'=0$ and 2 transitions at higher $J$.
The observed decrease of $f$-value with $J$ is then consistent with an $\Omega=1$ upper state.

Molecular parameters fitted to the observed transitions to $C(14)$ $\Omega=1$ levels are given in Table~\ref{tab:fitted_constants}, with an assumed spin-orbit constant of $A=\cm{-14}$ in accordance with previous observations of this and nearby $C(v)$ levels for other isotopologues.\cite{Lewis2008a}

A summation of $f$-values for all observed $C(v=14,\Omega=1)\leftarrow X(0)$ transitions gives a value of \np{3.6e-6}.
Assuming $C(14)\leftarrow X(0)$ borrows intensity from a remote ${}^1\Pi_u\leftarrow X(0)$ transition, we computed $f$-values for transitions to all $C(v=14,\Omega)$ levels.
Then, a whole-band $f$-value extrapolated from the experimentally observed transitions is \np{9.0e-6}.

\subsection{$C\,{}^3\Pi_u(v=15)$}
\begin{figure*}
  \centering
  \includegraphics{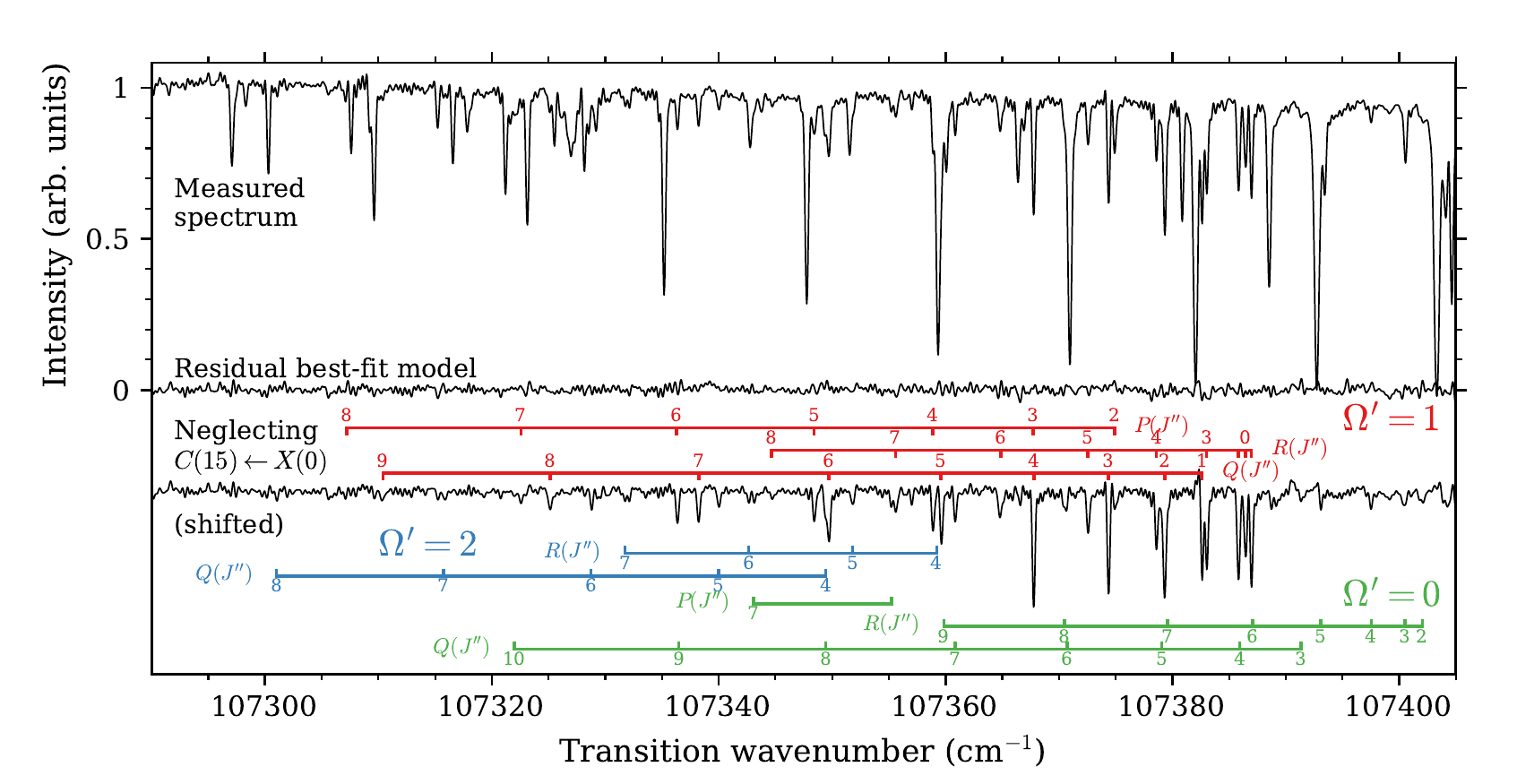}
  \caption{Spectrum showing $C(15)\leftarrow X(0)$ absorption, overlapping \ce{{}^{14}N^{15}N} transitions, and contamination from \ce{^{14}N2}, \ce{^{15}N2}, and \ce{H2}. The residual error after modelling this spectrum is shown, and for a model neglecting $C(15)\leftarrow X(0)$, to highlight its rotational structure.}
  \label{fig:C15_spectrum}
\end{figure*}
\begin{figure}
  \centering
  \includegraphics{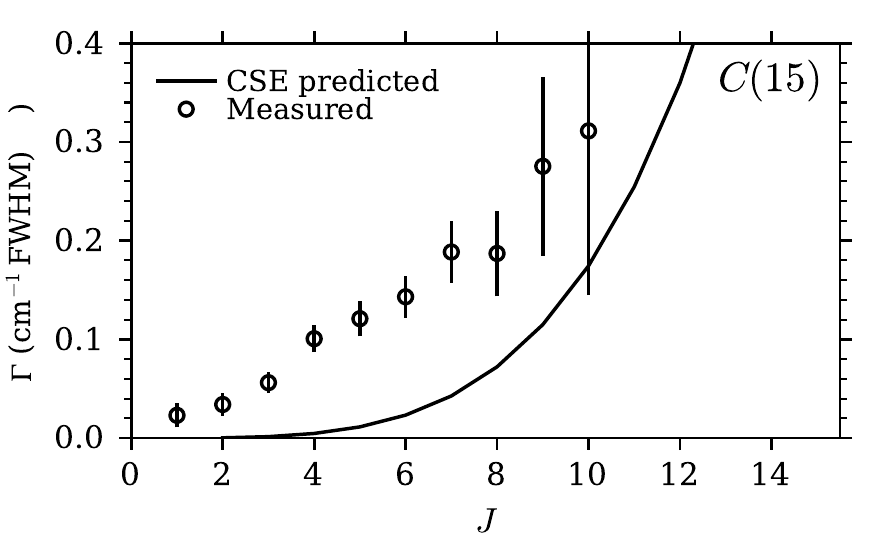}
  \caption{Rotational variation of predissociation linewidth for $C(15)$. Symbols: Experimental, this work. Curve: CSE computations using the
model of Ref.~\onlinecite{Lewis2008a}.}
  \label{fig:C15_widths}
\end{figure}

The $C(15)\leftarrow X(0)$ band has not been observed previously for any isotopologue and provides a test of the CSE model when interpolating both $v$ and reduced mass.
The upper trace in Fig.~\ref{fig:C15_spectrum} shows the measured absorption spectrum of $C(15)\leftarrow X(0)$, together with lines attributable to $b(9)\leftarrow X(0)$ and $o_3(1)\leftarrow X(0)$, weak contamination from the \ce{{}^{14}N_2} and \ce{{}^{15}N_2} bands $b'(5)\leftarrow X(0)$ and $b(9)\leftarrow X(0)$, and a few \ce{H2} contaminant lines.

The $C(15)$ level was modelled in tandem with $b(9)$ and a spin-orbit mixing between them, with all model parameters listed in Table~\ref{tab:fitted_constants}.
The resultant $C(15)\leftarrow X(0)$ lines are strongest for transitions to the upper state $\Omega'=1$ levels, but with weaker transitions to $\Omega'=0$ and $\Omega'=2$ also evident in the spectrum.
The best-fit model residual is plotted in Fig.~\ref{fig:C15_spectrum} as well as a residual neglecting all $C(15)\leftarrow X(0)$ absorption, to distinguish this from overlapping absorption.
The observation of all $C(15)$ $\Omega$-levels allows for a determination of both diagonal spin-orbit and spin-spin parameters, $A$ and $\lambda$, respectively.

The two-level model adopted here reproduces the wavenumbers and strengths of the $C(15)\leftarrow X(0)$ lines to experimental precision, but does not contain the necessary physics to constrain their predissociation broadening.
We fitted the latter independently while assuming $\Omega$- and $e/f$-parity-independence, with the resulting $J$-dependence plotted in Fig.~\ref{fig:C15_widths}.
These fitted widths are principally constrained by the strong $\Omega'=1$ transitions.

\subsection{$C\,{}^3\Pi_u(v=16)$}

\begin{figure}
  \centering
  \includegraphics{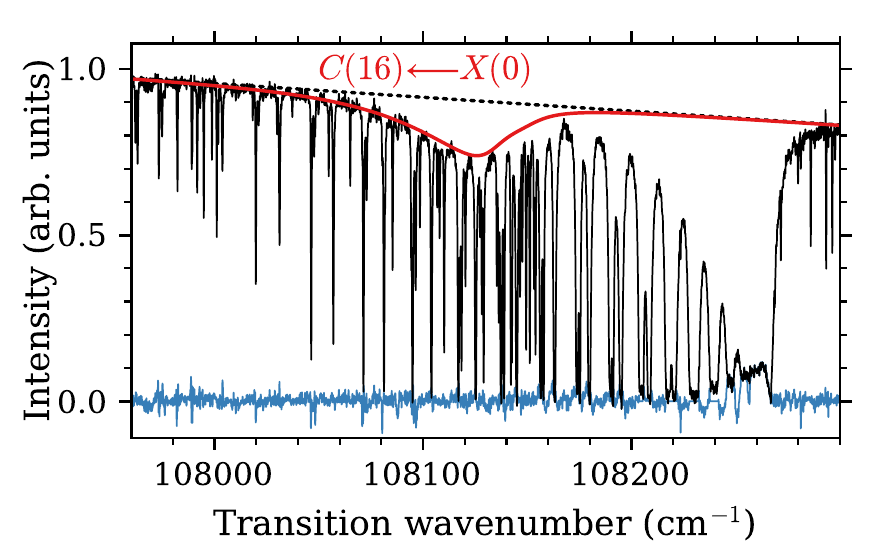}
  \caption{\emph{Black curve:} Absorption spectrum showing $C(16)\leftarrow X(0)$ and $b(10)\leftarrow X(0)$. \emph{Red curve:} $C(16)\leftarrow X(0)$ computed using a two-level deperturbation model. \emph{Blue curve:} Residual error of the full deperturbation model. Saturated $b(10)\leftarrow X(0)$ lines  are excluded from this residual. \emph{Dotted line:} Background synchrotron radiation intensity.}
  \label{fig:C16_spectrum}
\end{figure}

A broad absorption feature shown in Fig.~\ref{fig:C16_spectrum} at \cm{108120} is assigned to $C(16)\leftarrow X(0)$, based on the CSE model prediction and the known location of this band in \ce{{}^{14}N_2}.\cite{Lewis2008a}
The saturated bandhead of $b(10)\leftarrow X(0)$ occurs at \cm{108260}, showing some phase-error distortion, and higher-$J''$ lines also overlap $C(16)\leftarrow X(0)$.
Some further weak or intermediate strength lines are due to contamination from \ce{{}^{15}N_2}.
The \ce{{}^{14}N^{15}N} spectrum was treated with a two-level model including $C(16)$, $b(10)$, and a mutual spin-orbit interaction.
The strength of this mixing controls the intensity borrowing of $C(16)\leftarrow X(0)$ and how much of the predissociation broadening of $C(16)$ is transferred to $b(10)$.
The spin-orbit constant of $C(16)$ was fixed the near value computed by Ndome {\em et al.},\cite{Ndome2008} $A=-\cm{14}$, and assumptions of zero centrifugal distortion and $J$-independent line broadening made.
The absorption strength and profile of $C(16)\leftarrow X(0)$ is quite well fitted, as shown in Fig.~\ref{fig:C16_spectrum}, as is the line broadening of $b(10)\leftarrow X(0)$.
All fitted constants are given in Table~\ref{tab:fitted_constants} where uncertainties of the $T$, $B$, and $\Gamma$ parameters for $C(16)$ and the spin-orbit mixing parameter $\xi$ have been estimated manually with trial models.

\subsection{$C\,{}^3\Pi_u(v=21)$}
\begin{figure}
  \centering
  \includegraphics{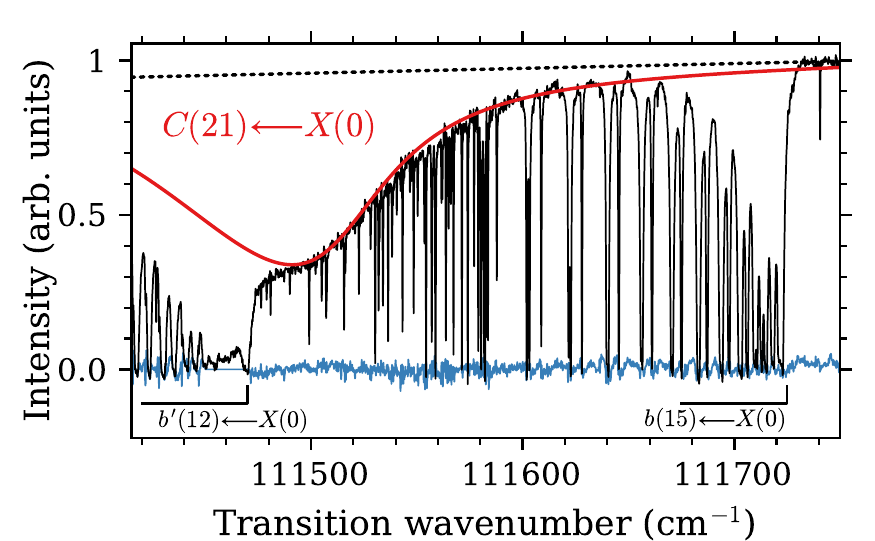}
  \caption{\emph{Black curve:} Absorption spectrum showing broad $C(21)\leftarrow X(0)$ feature, embedded under strong singlet bands, with
additional overlapping narrow lines due to \ce{{}^{14}N_2}. \emph{Red curve:} Fitted model $C(21)\leftarrow X(0)$ contribution.
\emph{Blue curve:} Residual error after modelling the spectrum. \emph{Dotted line:} Background synchrotron radiation intensity.} 
  \label{fig:C21_spectrum}
\end{figure}
A broad absorption feature shown in Fig.~\ref{fig:C21_spectrum} is assigned to $C(21)\leftarrow X(0)$ based on its close coincidence with the CSE-predicted band origin.
There is strong absorption due to $b'(12)\leftarrow X(0)$ obscuring much of the observed forbidden band and only two molecular parameters for $C(21)$, $T$ and $\Gamma$, could be fitted using the experimental spectrum.
For this fit, a spin-orbit constant of $A=\cm{-18}$ was assumed, following the values predicted for high $C(v)$ levels by Ndome {\em et al.},\cite{Ndome2008} and a rotational constant $B=\cm{1}$ was fixed near the predicted value of the CSE model.

\subsection{$G\,{}^3\Pi_u(v=0)$}
\label{sec:G00}
\begin{figure*}
  \centering
  \includegraphics{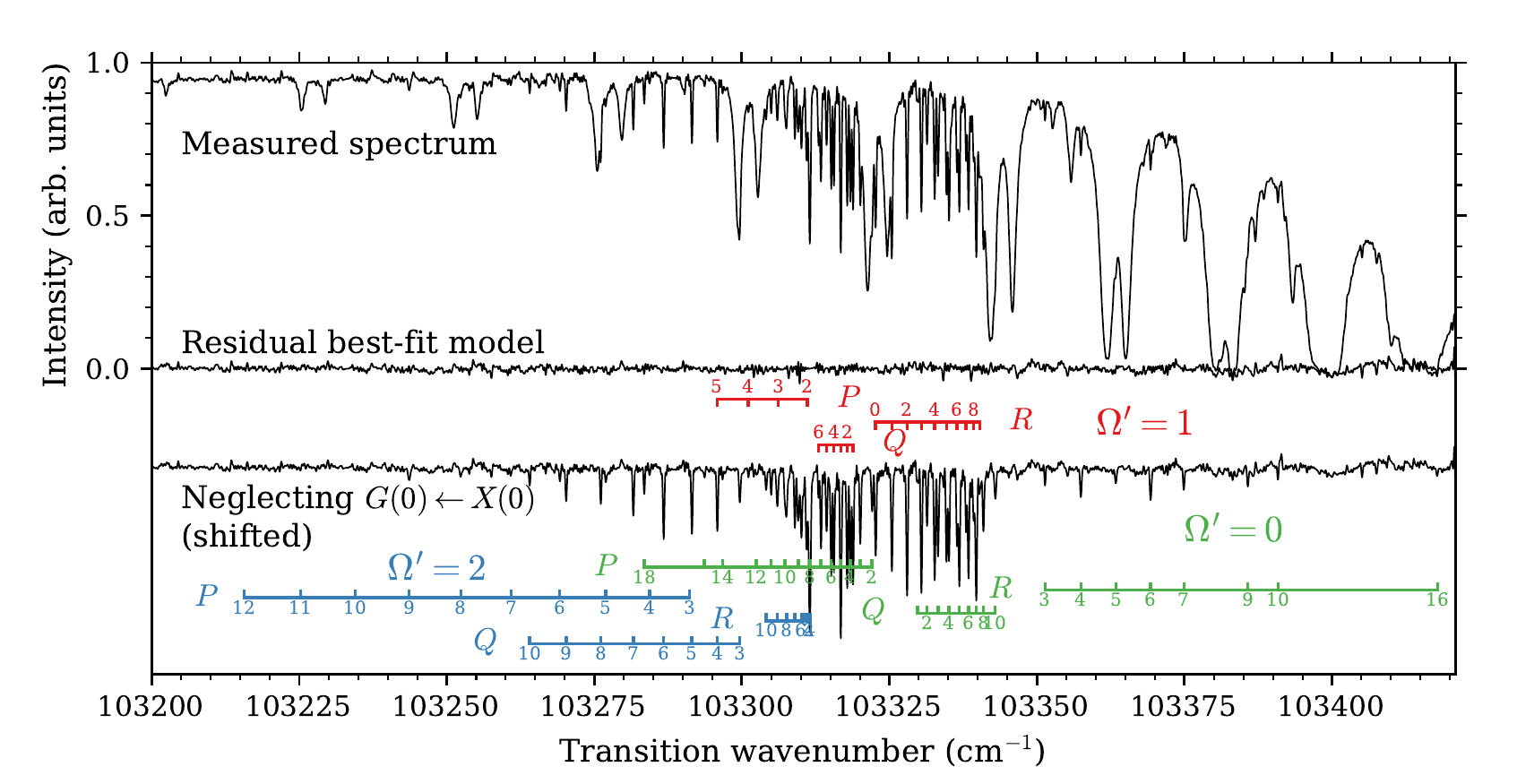}
  \caption{Spectrum and line assignments showing $G(0)\leftarrow X(0)$.  Saturated and broadened lines are due to $b(4)\leftarrow X(0)$ and contaminating \ce{H2}. The residual error after modelling this spectrum is shown, and for a model neglecting $G(0)\leftarrow X(0)$, to highlight its rotational structure.}
  \label{fig:G00_spectrum}
\end{figure*}
\begin{figure}
  \centering
  \includegraphics[width=\linewidth]{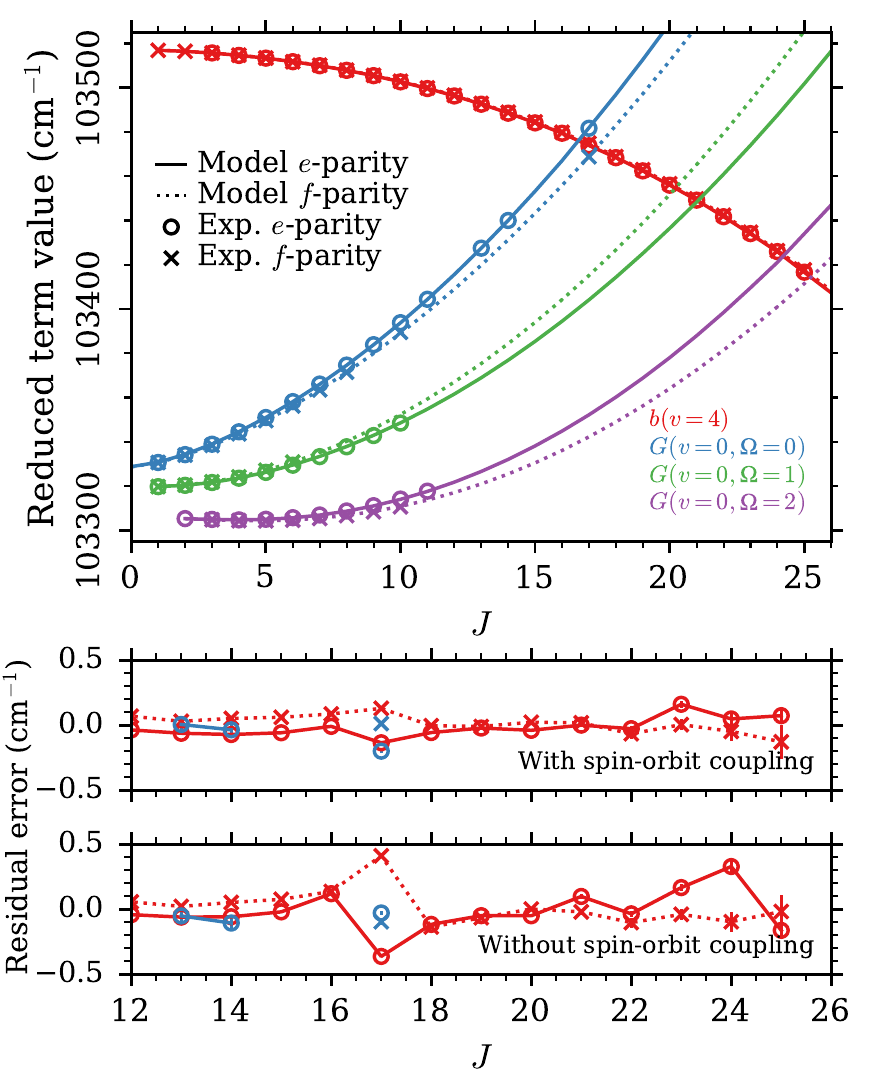}
  \caption{\emph{Upper figure}: Experimental term values of $G(0)$ and $b(4)$, reduced by the subtraction of $1.5\times J(J+1)$ are shown as symbols and term values computed from an effective Hamiltonian model as curves. \emph{Lower figures:}  Residual errors of effective Hamiltonian models, with and without the inclusion of $b(4) \sim G(0)$ spin-orbit coupling.}
  \label{fig:G00_b04_term_values}
\end{figure}

A spectrum of $G(0)\leftarrow X(0)$ is plotted in Fig.~\ref{fig:G00_spectrum} and line parameters are fitted to all nine of the expected rotational branches, with
a common linewidth found for the entire spectrum.
This band borrows strength principally through the interaction between $G(0)$ and $b(4)$.
We simultaneously fit a low-pressure spectrum of $b(4)$ measured by Heays {\em et al.}\cite{Heays2011} to properly characterize this interaction.

A near-degeneracy of $G(0)$ $\Omega=0$ $e/f$-parity levels with $b(4)$ occurs at $J=17$, leading to a perturbation of $b(4)$ energies and the appearance of extra $Q(17)$, $R(16)$ and $P(18)$ lines attributable to $G(0)\leftarrow X(0)$.
Similar $b(4)$ energy shifts also occur where it crosses the $\Omega=1$ and 2 levels of $G(0)$ near $J=21$ and 25, respectively, but without the occurrence of observable extra lines in our spectra.

The experimental energy levels of $G(0)$ and $b(4)$ are plotted in Fig.~\ref{fig:G00_b04_term_values}, along with term values computed from an effective Hamiltonian model which includes these states and their spin-orbit interaction.
The fitted parameters of this model are given in Table~\ref{tab:fitted_constants} and its residual error is shown in Fig.~\ref{fig:G00_b04_term_values}.
A trial model neglecting the spin-orbit interaction of $b(4)$ and $G(0)$ fails to account for the observed $b(4)$ levels at all three crossing points, with residual errors up to \cm{0.5} demonstrated in Fig.~\ref{fig:G00_b04_term_values}.
Then, the $\Omega=1$ and 2 sublevels of $G(0)$ between their observed levels and $b(4)$ crossings are likely to be well-interpolated by the two-level Hamiltonian model.

Figure~\ref{fig:G00_spectrum} provides an instructive example of the ${^3\Pi_u} \sim {^1\Pi_u}$ mixing complexities discussed at the end of Sec.~\ref{sec:analysis}.
There is an
extremely small $G(0)\sim b(4)$ mixing, yet it is large enough to provide all of the observed $G(0)\leftarrow X(0)$ oscillator strength. However, the same mixing has
essentially no effect on the unmixed $G(0)$ and $b(4)$ level widths, 0.11~cm$^{-1}$ FWHM and 1.46~cm$^{-1}$ FWHM,\cite{Heays2011} respectively.
As evidenced by the CSE model of Lewis {\em et al.},\cite{Lewis2008b} employed here, the $G(0)$ width is explicable using only a manifold of the coupled $F$, $G$, $C$, and 
$C^{\prime}$ $^3\Pi_u$ states. Similarly, the widths of the low $b(v)$ levels, including $b(4)$, are explicable using a CSE model containing only the $b$, $c$, $o$ 
$^1\Pi_u$ manifold, together with the $C$ and dissociative $C^{\prime}$ $^3\Pi_u$ states.\cite{Lewis2005a} Remarkably, despite all predissociation in this region being
derived from the $^3\Pi_u$ route, the $G(0)$ level plays no measurable part in the predissociation of $b(4)$.

\subsection{$G\,{}^3\Pi_u(v=1)$}

\begin{figure}
  \centering
  \includegraphics{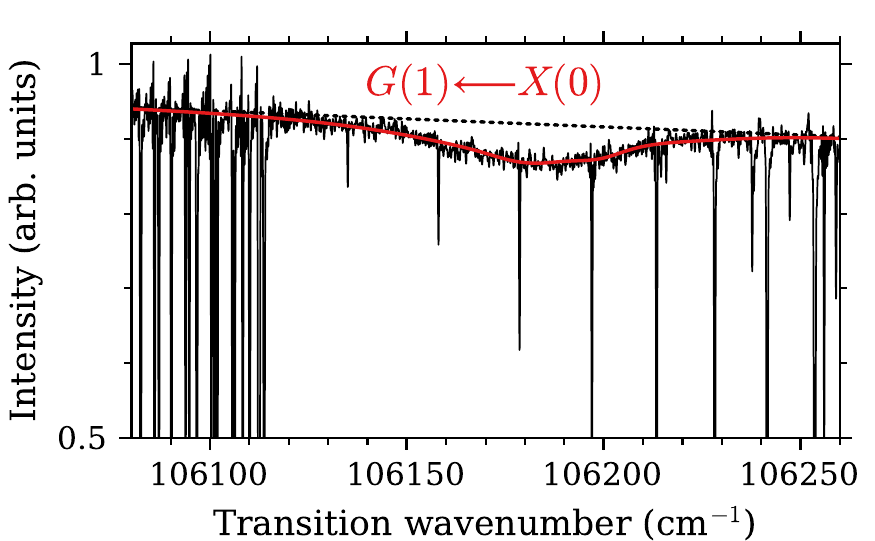}
  \caption{
\emph{Black curve:} Absorption spectrum showing $G(1)\leftarrow X(0)$.  Narrow overlapping features are due to $c'_4(1)\leftarrow X(0)$ from \ce{{}^{14}N{}^{15}N} and contaminating $b(7)\leftarrow X(0)$ lines from \ce{{}^{14}N_2}.
\emph{Red curve:} Modelled $G(1)\leftarrow X(0)$ absorption.
\emph{Dotted line:} The frequency-dependent synchrotron radiation intensity.
  }
  \label{fig:G01_spectrum}
\end{figure}

A broad absorption feature, shown in Fig.~\ref{fig:G01_spectrum}, is observed in the VUV-FT spectrum near \cm{106180},
close to the position of a similar broad feature in the absorption spectrum of cooled \ce{{}^{14}N_2} which has been assigned as $G(1)\leftarrow X(0)$.\cite{Lewis2008a}
Principally on the basis of intensity considerations, we assign this feature to $G(1)\leftarrow X(0)$ in \ce{{}^{14}N{}^{15}N}, but note that this assignment may not be quite as robust as others in this work.
A band model has been constructed in order to constrain the term origin and predissociation linewidth of $G(1)$,
assuming a fixed spin-orbit constant, $A=\cm{-8}$, in accord with the value deduced for $G(0)$, as well as zero centrifugal distortion, and $J$-independent predissociation broadening.
The intensity of $G(1)\leftarrow X(0)$ is due to spin-orbit interaction with one or more nearby ${}^1\Pi_u$ levels and this is approximated with a two-level model comprising $G(1)$ spin-orbit mixed with $b(7)$ (term origin of \cm{106045}).
The fitted term origin, rotational constant, and predissociation linewidth for $G(1)$, with uncertainties subjectively estimated following the consideration of a range of parameter values, are given in Table~\ref{tab:fitted_constants}, with the best-fitting model spectrum indicated in Fig.~\ref{fig:G01_spectrum}.

\subsection{$G\,{}^3\Pi_u(v=4)$}
\begin{figure}
  \centering
  \includegraphics{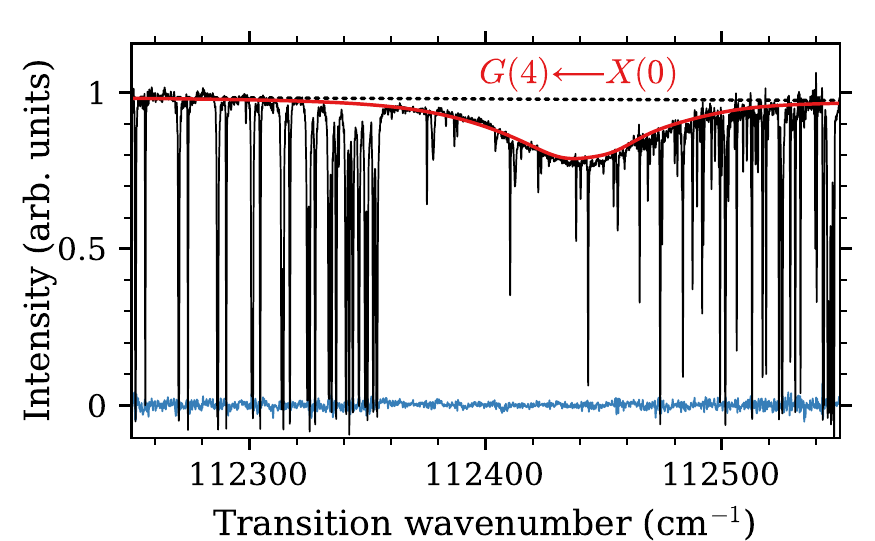}
  \caption{\emph{Black curve:} Absorption spectrum showing $G(4)\leftarrow X(0)$.  Narrow overlapping features are due to $b(16)\leftarrow X(0)$ absorption from \ce{{}^{14}N{}^{15}N} and $b(7)\leftarrow X(0)$ absorption from \ce{{}^{14}N_2} as well as \ce{H2} contamination. \emph{Red curve:} Modelled $G(4)\leftarrow X(0)$ absorption. \emph{Blue trace:} Residual error of the deperturbation model. \emph{Dotted line:} The frequency-dependent synchrotron radiation intensity.}
  \label{fig:G04_spectrum}
\end{figure}

A diffuse feature centred at \cm{112450} is assigned to $G(4)\leftarrow X(0)$ absorption, since it occurs in the region of the CSE-predicted  wavenumber for this band.
Its spectrum is plotted in Fig.~\ref{fig:G04_spectrum} together with an effective Hamiltonian simulation.
In this model, the intensity-borrowing of $G(4)\leftarrow X(0)$ is attributed to the spin-orbit coupling of $G(4)$ with a single remote ${}^1\Pi_u$ level, approximating the multilevel interactions mixing ${}^1\Pi_u$ and singlet manifolds.

We assume a spin-orbit constant $A=\cm{-8}$, in common with resolved observations of other $G(v)$ levels and isotopologues, then finding a rotational constant $B=1.76(1)$\cm{}, consistent with expectation for a $G(v)$ or $F(v)$ Rydberg level.
An equally good fit to the experimental data is obtained by assuming that the observed absorption corresponds to a an $F(v)$ level with $A=\cm{-20}$.
This assignment, however, requires a large deviation from any reasonable energy extrapolation of $F(v)$ levels, known up to $v=3$ in both \ce{{}^{14}N_2} and \ce{{}^{15}N_2}.\cite{Lewis2008b}
We rule out an assignment to a high-$v$ $C$-state level because the required rotational constant of $B\simeq \cm{1}$ is incompatible with a good fit to the spectrum in Fig.~\ref{fig:G04_spectrum}.

\subsection{$F\,{}^3\Pi_u(v=0)$}
\begin{figure*}
  \centering
  \includegraphics{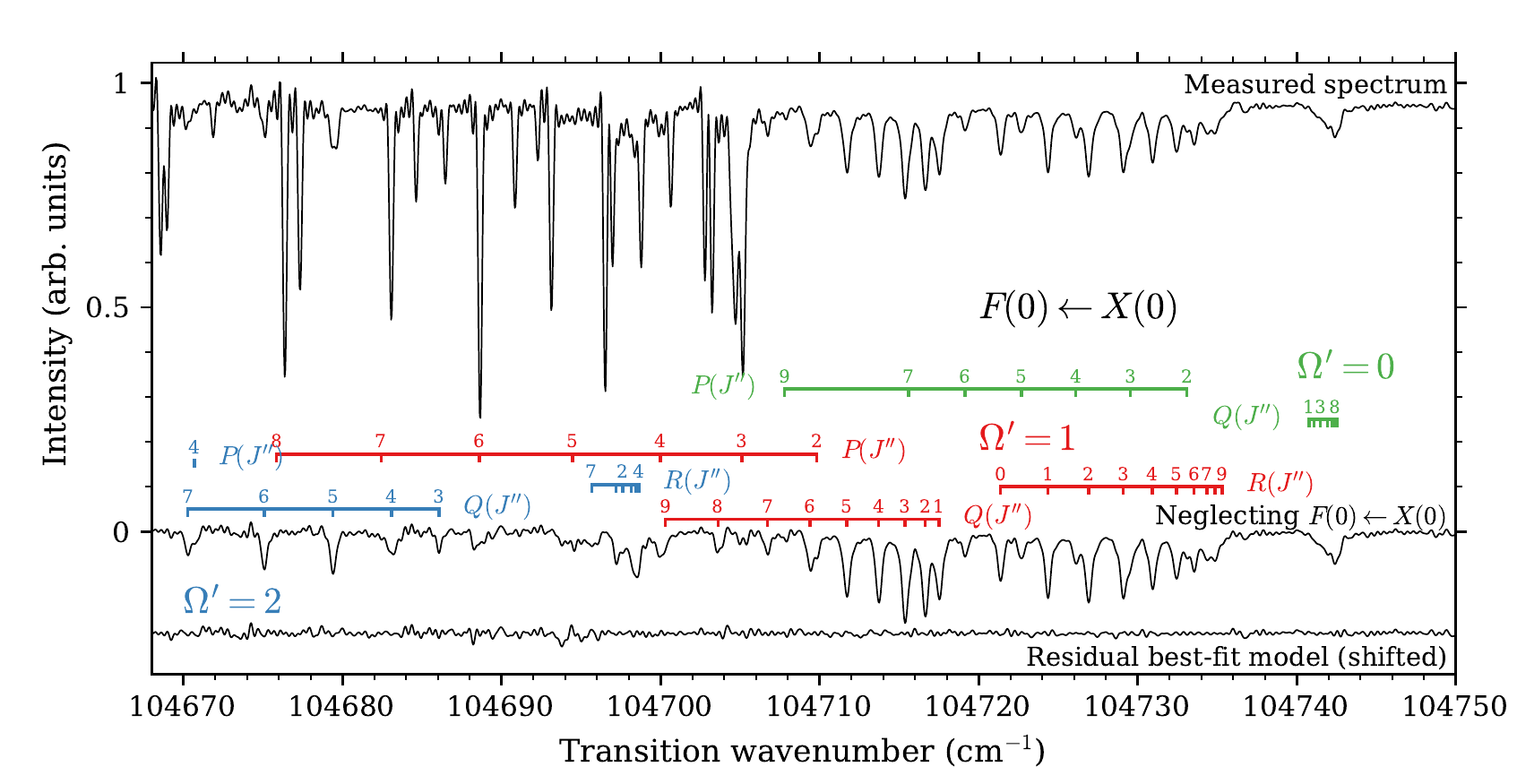}
  \caption{Spectrum showing $F(0)\leftarrow X(0)$ absorption and overlapping \ce{{}^{14}N_2} and \ce{H2} contamination. The residual error after modelling this spectrum is shown, and for a model neglecting $F(0)\leftarrow X(0)$, to highlight its rotational structure.}
  \label{fig:F00_spectrum}
\end{figure*}
Eight rotational branches accessing levels with $J'$ up to 10 are observed near \cm{104700} and attributed to $F(0)\leftarrow X(0)$ absorption, with intensity borrowed from the nearby $b(5)\leftarrow X(0)$ band.
The observed transitions, shown in Fig.~\ref{fig:F00_spectrum}, were used to determine the molecular parameters for $F(0)$ listed in Table~\ref{tab:fitted_constants} and are well described by rotation-independent predissociation broadening.

\section{Comparison with the CSE model}
\label{sec:CSE model comparison}
\begin{figure}
  \centering
  \includegraphics{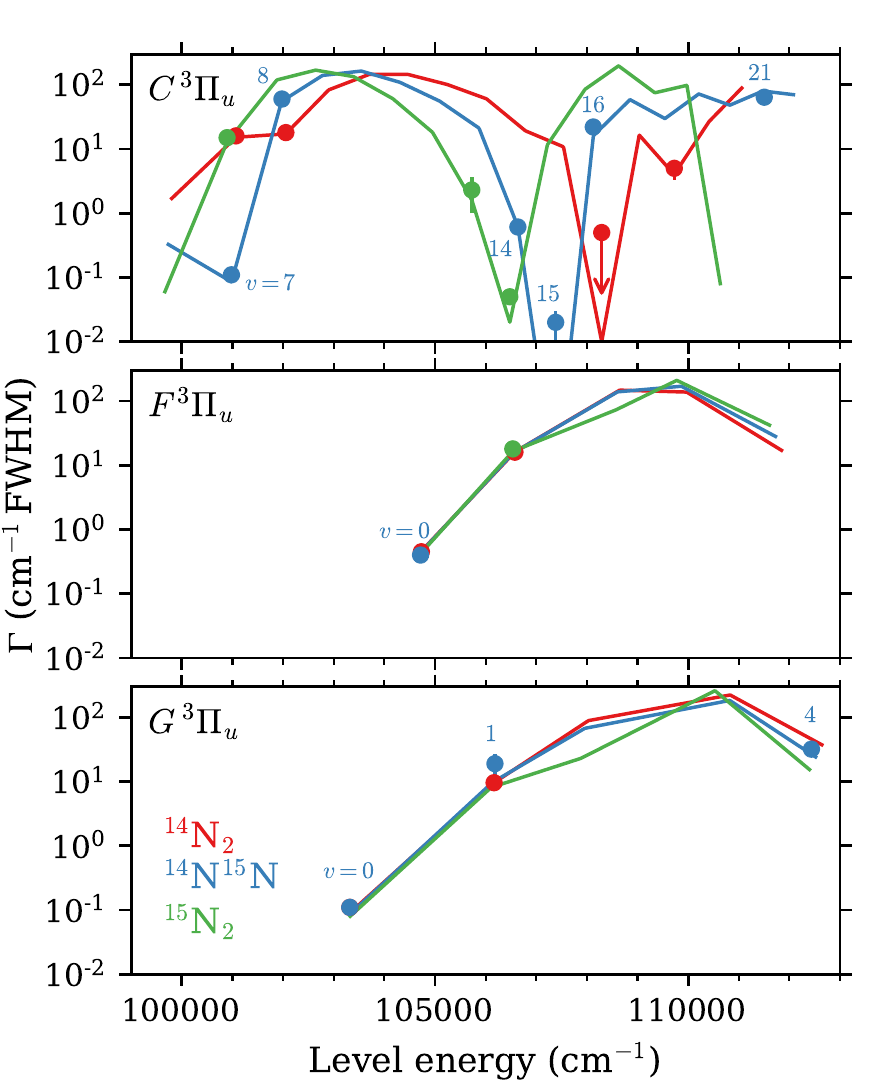}
  \caption{Comparison between experimental and CSE-modelled ${}^3\Pi_u$ predissociation linewidths for isotopologues of N$_2$.
Symbols: Experimental values for \ce{{}^{14}N{}^{15}N} (present work), \ce{{}^{14}N2} and \ce{{}^{15}N2} (Ref.~\onlinecite{Lewis2008b}).
Line vertices: CSE computations using the model of Ref.~\onlinecite{Lewis2008a}.}
  \label{fig:widths}
\end{figure}

A comparison between the experimental and CSE-computed ${}^3\Pi_u$ molecular parameters and linewidths for \ce{{}^{14}N{}^{15}N} is made in Table~\ref{tab:fitted_constants}.
Most term origins are in agreement within a few cm$^{-1}$ apart from the $G(1)$ and $G(4)$ levels,
where the energy discrepancies are significantly in excess of the corresponding predissociation linewidths. Strictly speaking, the $G(4)$ region is above the energy range of applicability of the present CSE model and the new experimental data here should enable future model improvement. 
While the experimental and model terms are in excellent agreement for $C(14)$ and $C(16)$, the $\sim 4$~cm$^{-1}$ difference for $C(15)$
is surprisingly large for such a narrow resonance, especially since the model has been optimized using precisely-known narrow resonances, $C(14)$ for
$^{14}$N$_2$ and $C(16)$ for $^{15}$N$_2$.\cite{Lewis2008a}
Clearly, the present observation of $C(15)$ in \ce{{}^{14}N{}^{15}N} should enable further constraint of the CSE model in this energy region.
Computed rotational constants are within 1\% of the experimental values, or within the experimental uncertainty where this is larger, apart from $C(14)$, marginally, and $C(16)$. The difference for $C(16)$ probably reflects the difficulty in fitting rotational constants  
and determining their uncertainties using diffuse spectra.

A comparison between experimental and modelled predissociation widths for all isotopologues is presented in Fig.~\ref{fig:widths}, with \ce{{}^{14}N2} and \ce{{}^{15}N2} model and experimental widths taken from the collation of Lewis {\em et al.}\cite{Lewis2008b}
Collectively, experimental studies of the various isotopologues have observed all levels with energies below \cm{110000} and widths below   \cmFWHM{20}, respectively, with two broader bands, $C(8)$ and $C(21)$, identified here.
The difficulty of studying weak, diffuse bands amidst the dense allowed singlet spectrum of \ce{N2} explains this selection.

The multiple order-of-magnitude oscillation of $C$-state predissociation linewidths with vibrational level, evident in Fig.~\ref{fig:widths}, is somewhat reminiscent of predissociation induced by an outer-limb-crossing repulsive state,\cite{Lefebvre} a role played in this case by $C'\,{}^3\Pi_u$.
However, the present case is much more complicated than the textbook perturbative crossing, due to the relatively large $C \sim C^{\prime}$ interaction, configurational mixing in the $C$ state which leads to the unusual shape of its potential-energy curve, and the multistate nature of the predissociation.
In particular, Rydberg-valence and Rydberg-Rydberg interactions mixing the predissociated ${^3\Pi_u}$ bound states have the potential to produce interference effects in the width matrix elements which can profoundly affect level widths, especially in the energy region above $G(0)$ and $F(0)$.  
The experimental level widths in Fig.~\ref{fig:widths} are very well reproduced by the CSE model, including where a reduced-mass interpolation is made to \ce{{}^{14}N{}^{15}N} levels.
In particular, deep width minima for $C\,{}^3\Pi_u$ are predicted correctly at $v=7$ and $v=15$.

The computed widths in Table~\ref{tab:fitted_constants} and Fig.~\ref{fig:widths} are for virtual $J=0$ levels.
We observe $C(7)$ and $C(15)$ in \ce{{}^{14}N{}^{15}N} to have $J$-dependent linewidths and these are compared with an excited-$J$ CSE calculation in Figs.~\ref{fig:C07 widths} and \ref{fig:C15_widths}, respectively.
The monotonically increasing linewidths of $C(7)$ are well predicted by the CSE model, no doubt because this level was found by Lewis {\em et al.}\cite{Lewis2008b} in both \ce{{}^{14}N2} and \ce{{}^{15}N2}, and was included in their CSE-model fitting.\cite{Lewis2008a}
Furthermore, the monotonic increase with $J$ of the $C(15)$ widths is also predicted by the model, albeit with a lesser slope.
That the CSE widths for these two levels are a little lower than the experimental values is unsurprising, given that the CSE model includes only the strong electrostatic interactions, with no spin-orbit or rotational couplings. 
The effects of this approximation will be most noticeable for the narrowest levels, where the electrostatic contribution will be very close to zero.

The predissociation widths for the lowest Rydberg levels, $G(0)$ and $F(0)$, are also very well predicted by the model. However, 
the CSE model of Lewis {\em et al.},\cite{Lewis2008b} which is employed here to describe the ${}^3\Pi_u$ states, is most limited in the energy region of $F(2)$, $G(2)$, and above, since it is based on experimental data having at least 50~cm$^{-1}$ energy uncertainty in that region. 
Furthermore, the model does not include the the next-highest $^3\Pi_u$ valence state, $III\,{^3\Pi_u}$,\cite{Lewis2008a} which will definitely affect
levels above $\sim 112\,000$~cm$^{-1}$.
As stated above, the 
present experimental data, particularly for $G(4)$, will improve that situation, allowing refinement of the model potential-energy curves and electrostatic couplings.
To best accomplish that future task, it will be preferable to include, not only direct ${}^3\Pi_u$ data for all isotopologues, but also indirect experimental constraints on further unobserved ${}^3\Pi_u$ levels due to their spin-orbit interaction with optically-excited ${}^1\Pi_u$ and ${}^1\Sigma^+_u$ states which show resultant level broadening.\cite{Heays2011}
The result will be a multichannel model including triplet and singlet states coupled by electrostatic, spin-orbit, and rotational interactions.

\section{Conclusions}

Ten spin-forbidden vacuum-ultraviolet bands of \ce{{}^{14}N{}^{15}N} have been observed and spectroscopically analysed in a direct photoabsorption experiment, specifically, the $v=7$, 8, 14, 15, 16, and 21 levels of $C\,{}^3\Pi_u$, $v=0$, 1, and 4 levels of $G\,{}^3\Pi_u$, and $v=0$ level of $F\,{}^3\Pi_u$.
Assignment of the observed bands was made with the aid of wavenumbers and linewidths predicted using a coupled Schr\"odinger-equation model of electrostatically-coupled ${}^3\Pi_u$ states, originally designed to reproduce a database of \ce{{}^{14}N_2} and \ce{{}^{15}N_2} ${}^3\Pi_u$ levels.
In turn, the new \ce{{}^{14}N{}^{15}N} data provide a convincing demonstration of the predictive power of the CSE model in terms of mass and energy interpolation.
When combined with our previous experimental study of \ce{{}^{14}N{}^{15}N} singlet states, the present work provides a sound basis for a general refinement and extension of the CSE model of N$_2$ spectroscopy and predissociation.

\section*{Supplementary Material}

Supplementary material to this article contains a line-by-line listing of all modelled transitions frequencies, oscillator strengths, and linewidths fit to the experimental spectra.
For the rotationally blended bands that are modelled with effective Hamiltonians lines, these are listed as far as $J''\leq15$.
Also provided are excited-state rotational level energies and widths deduced from the observed transitions.

\section*{Acknowledgements}

This work has been carried out within the framework of the Dutch Astrochemistry Network funded by NWO. The authors are grateful to the staff at Soleil for their hospitality and operation of the facility under project number 20120023.
For portions of this work AH was funded by NASA Postdoctoral Program through the NASA Astrobiology Institute.
The authors thank Dr. E.~F.~van~Dishoeck for valuable discussions on N$_2$ predissociation dynamics, including their significance in isotopic fractionation.

\clearpage
\newpage




\begin{thebibliography}{99}
\bibitem{FN1}In this paper we have indicated the $u/g$ inversion symmetry quantum label even though \ce{{}^{14}N^{15}N} is a heteronuclear species, due to the smallness of the perturbation attributable to this symmetry breaking relative to our experimental precision.
\bibitem{FN4}In this work, the spectroscopic notation is usually abbreviated where this is unambiguous, e.g., $C\,{}^3\Pi_u(v'=7)\leftarrow X\,{}^1\Sigma^+_g(v''=0)$ to $C(7)\leftarrow X(0)$.
\bibitem{FN2}Radiative lifetimes are too long to lead to measurable broadening.
\bibitem{FN3}In this work, we use the term ``absorption spectrum'' in its general sense, throughout. Strictly, the model fitting to 
the experimental data is performed in the transmittance space.
\end{thebibliography}
\end{document}